%% file: tambyah_lee_badia_2025_arxiv.tex
\documentclass[oneside, reqno, 11pt, a4paper]{amsart}

\usepackage{microtype}
\AtBeginDocument{
	\DeclareSymbolFont{AMSb}{U}{msb}{m}{n}
	\DeclareSymbolFontAlphabet{\mathbb}{AMSb}
}

\usepackage[dvipsnames]{xcolor}
\usepackage{graphicx}
\usepackage{tikz}
\usepackage{subfig}
\usepackage{svg}

\usepackage[a4paper, pdftex, left=2cm, top=2cm, right=2cm, bottom=2cm]{geometry}
\usepackage[final]{pdfpages}
\usepackage{changepage}

\usepackage[foot]{amsaddr}

\usepackage{url}
\usepackage{hyperref}
\hypersetup{
	breaklinks=true,
	bookmarksopen=true,
	pdftitle={Entropy conservation for the thermal shallow water equations}, 
	pdfauthor={Santiago Badia, Tamara Tambyah, David Lee}, 
	pdfsubject={partial differential equations, thermal shallow water equations}, 
	pdfkeywords={thermal shallow water equations, compatible finite elements, entropy conservation, Poisson systems, Casimir conservation}, 
	colorlinks=true,
	linkcolor=black,
	citecolor=blue,
	filecolor=black,
	urlcolor=blue
}
\usepackage[capitalise]{cleveref}
\crefname{equation}{}{}
\crefrangeformat{equation}{(#3#1#4)--(#5#2#6)}

\usepackage{csquotes}
\usepackage[backend=biber, defernumbers=false, maxbibnames=99, style=numeric-comp, isbn=false, bibencoding=utf8, safeinputenc, url=false, doi=true, giveninits=true,sorting=none,natbib=true,uniquelist=false,maxcitenames=2]{biblatex}

\usepackage{float}
\usepackage{framed}
\usepackage{verbatim}
\usepackage{fancyvrb}
\usepackage{booktabs}
\usepackage{multirow}
\usepackage{algorithm}
\usepackage{algpseudocode}
\usepackage[inline]{enumitem}

\newtheorem{theorem}{Theorem}[section]

\newtheorem{proposition}[theorem]{Proposition}

\usepackage{soul}

\usepackage{amsmath, amsfonts, amssymb, amscd}
\usepackage{mathtools}
\usepackage[only, llbracket, rrbracket]{stmaryrd}
\allowdisplaybreaks

\DeclareMathOperator{\sign}{sign}

\usepackage{acronym}

\title[Entropy conservation]{
Energy and entropy conserving compatible finite elements with upwinding for the thermal shallow water equations
}
\date{\today}

\author[1]{Tamara A.  Tambyah$^{1\ast}$}
\address{$^1$School of Mathematics\\Monash University\\Clayton\\Victoria 3800\\Australia }
\email{tamara.tambyah@monash.edu}

\author[2]{David Lee$^{2}$}
\email{david.lee@bom.gov.au}
\address{ $^{2}$Bureau of Meteorology\\Melbourne\\Australia}

\author[3]{Santiago Badia$^{1}$}
\email{santiago.badia@monash.edu}
 
\thanks{$^\ast$Corresponding author}

\newcommand{\reviewone}[1]{{\color{black}#1}}
\newcommand{\reviewtwo}[1]{{\color{black}#1}}
 
\newcommand{\reviewonee}[1]{{\color{black}#1}}
\begin{document}

\begin{abstract}
  \input{abstract.tex}

\end{abstract}

\maketitle

\input{introduction.tex}

\input{tsw.tex}

\input{fem.tex}

\input{results.tex}

\input{conclusion.tex}

\section*{Acknowledgements}
This research is supported by the Commonwealth of Australia as represented by the Defence Science and Technology Group of the Department of Defence.
This research is also funded by the Australian Government through the Australian Research Council (project numbers DP210103092 and DP220103160). 
This work is supported by computational resources provided by the Australian Government through NCI under the National Computational Merit Allocation Scheme (NCMAS), ANU Merit Allocation Scheme, and the Monash-NCI partnership program.
The authors acknowledge the two anonymous reviewers, whose comments helped to improve the quality of this article.

\subsection*{Author contribution}
All authors contributed equally to the design of the study. 
Tamara A. Tambyah performed numerical simulations and drafted the article. All authors gave approval for publication. 

\subsection*{Declaration of competing interest}
The authors declare they have no competing interests. 

\subsection*{Data availability}
This articles does not contain any additional data. 
The source code for this study is available on Zenodo \cite{Zenodo}.

\printbibliography

\end{document}

%% file: abstract.tex
In this work, we develop a new compatible finite element formulation of the thermal shallow water equations that conserves energy and mathematical entropies given by buoyancy--related quadratic tracer variances. 
Our approach relies on restating the governing equations to enable discontinuous approximations of thermodynamic variables and a variational continuous time integration. A key novelty is the inclusion of centred and upwinded fluxes. The proposed semi-discrete system conserves discrete entropy for centred fluxes, monotonically damps entropy for upwinded fluxes, and conserves energy. 
The fully discrete scheme \reviewone{preserves} entropy conservation at the continuous level.  
The ability of a new linearised Jacobian, which accounts for both centred and upwinded fluxes, to capture large variations in buoyancy and  simulate thermally unstable flows for long periods of time is demonstrated for two different transient case studies. 
The first involves a thermogeostrophic instability where including upwinded  fluxes is shown to suppress spurious oscillations while successfully conserving energy and monotonically damping entropy. 
The second is a double vortex where a constrained fully discrete formulation is shown to achieve exact entropy conservation in time.

%% file: introduction.tex
\section{Introduction}
\label{sec:introduction}
 
Mathematical entropies, or \textit{entropy}, are convex functionals arising from a positive definite Hessian \cite[Chapter 3]{Fisher2013,Leveque1992} that correspond to quadratic buoyancy--related  tracer invariants for the thermal shallow water equations \cite{Ricardo2023,Ricardo2024-dg}.  
The thermal shallow water equations are a useful stepping stone from simpler atmospheric systems, like the rotating shallow water equations, to the full three-dimensional compressible Euler equations
typically used to describe atmospheric dynamics in operational weather models \cite{Maynard2020,Adams2019,Lee2021}.
The thermal shallow equations are analogous to the compressible Euler equations  with an identical Poisson bracket and entropy, and include a thermodynamic scalar quantity reflective of temperature \cite{Eldred2019,Ricardo2023,Ricardo2024-dg}.
Such thermodynamic quantities are not included in the rotating shallow water equations \cite{Cotter2014,BauerCotter2018,Mcrae2013,Wimmer2020}, meaning \reviewone{this specific model }does not capture the effect of thermodynamic transport on the horizontal pressure gradient \cite{Kurganov2021}. 
In this study, we develop a novel finite element approximation of the thermal shallow water equations, \reviewone{ and prove conservation of energy and entropy at the semi-discrete level.
}
Large scale simulations show the resulting scheme is conservative,  captures turbulent dynamics, and can stably simulate non-linear flow  over long time scales where a mature turbulent state is reached.

The desired properties of state-of-the-art numerical solvers for atmospheric models, discussed by \citet{Staniforth2012}, emphasise conserving  system invariants over long time scales 
 \cite{Gibson2019,Thuburn2008}. 
Energy is a key invariant that can be  conserved  by exploiting the non-canonical Hamiltonian form of the governing equations at the semi-discrete level \cite{Shepherd1990,Salmon1998}. 
Discrete entropy conservation improves model stability of hyperbolic systems that involve thermodynamic quantities by bounding unstable growth associated with  grid scale variance  \cite{Ricardo2023,Ricardo2024-dg}. 
Recent studies achieve semi-discrete entropy conservation by rewriting the equations of motion to allow for discontinuous approximations of thermodynamic variables \cite{Ricardo2023,Ricardo2024-dg}. 
In this study, we take a similar approach and restate the thermal shallow water equations to obtain semi-discrete entropy conservation under continuous time integration.

Entropy conservation requires preserving certain conformity requirements in space and time that may not be inherited by the numerical approximations.
A previous entropy conserving study uses a mixed finite element method in conjunction with Galerkin projections to enforce the necessary regularity of  buoyancy fluxes \cite{Ricardo2023}. Such continuity requirements are achieved in a different study via a discontinuous Galerkin method \cite{Ricardo2024-dg}. 
In the current study, we consider  a compatible finite element method that includes internal element boundary fluxes, cast in either a centred or upwinded form. 
The centred fluxes conserve energy and entropy, while the upwinded fluxes are proven to conserve energy and monotonically damp entropy. 
Care is taken to derive buoyancy terms that ensure entropy exchanges are balanced in both space and time for all forcing terms. In doing so, we ensure the only source of entropy conservation error is from temporal derivatives.
The semi-discrete formulation also
conserves total mass, buoyancy, vorticity, and supports compatible advection of buoyancy,  thus satisfying the criteria of a compatible  finite element discretisation for the thermal shallow water equations \cite{Eldred2019,Ricardo2024-dg}.

Constructing Poisson time integrators for non-canonical Hamiltonian systems often requires exploiting specific structures of the model problem \cite[Chapter VII.4]{Hairer2006}. 
The Poisson time integrator proposed by \citet{Cohen2011} conserves energy through exact temporal integration of the variational derivatives of the Hamiltonian  \cite{Eldred2019,BauerCotter2018,Lee2021,Lee2022,Wimmer2020,Cotter2014}. 
Quadratic invariants are also temporally preserved via the Poisson integrator \cite{Cohen2011}, while cubic invariants, which represent mathematical entropies in the form of tracer variances, are generally not.
Previous studies \cite{Ricardo2023,Ricardo2024-dg} regarding entropy conservation consider strong stability preserving time integrators \cite{Shu1988,Durran2010} for which entropy is not conserved exactly in time and energy variance is damped.  
In this work, we take a different approach and 
use a Poisson integrator \cite{Cohen2011} to construct a fully discrete scheme that conserves energy. 
Discrete buoyancy is represented as a linear polynomial in time such that discrete entropy is a cubic  polynomial in time, which is not temporally conserved pointwise by the chosen Poisson integrator \cite{Cohen2011}.    
We prove the loss in exact entropy conservation depends on the accuracy of the temporal approximation, and consequently propose  
a constrained formulation using Lagrange multipliers for which entropy is exactly conserved in time.

This article is structured as follows: in \cref{sec: tsw} we reformulate the thermal shallow water equations  at the continuous level and analyse the regularity requirements for continuous entropy conservation. 
Finite element approximations are derived in \cref{sec: discretisation,sec: temp discretisation}, where  conservation at the semi- and fully discrete levels is evaluated. 
In \cref{sec: results}, 
convergence of the new scheme under $h$-$p$ refinement is demonstrated using a steady thermogeostrophic balance test case \cite{Eldred2019}. 
By considering small perturbations from the solution at progressive time levels instead of the usual mean flow state \cite{Eldred2019,Lee2022}, we propose a new linearised Jacobian and quasi-Newton approach that shows robust convergence in the presence of large variations in buoyancy for both centred and upwinded fluxes.  Thus, we stably simulate thermally unstable flows 
to well evolved turbulent states, as demonstrated for transient case studies involving a thermogeostrophic instability \cite{Eldred2019,Gouzien2017,Kurganov2021,Zeitlin2018} and a  double vortex  \cite{Giorgetta2009,Eldred2019}. 
Conservation of invariants over long time scales is assessed, and  Lagrange multipliers are used to correct small temporal losses in entropy conservation for centred numerical fluxes.  
The inclusion of upwinded numerical fluxes is shown to suppress spurious oscillations while  conserving energy and monotonically damping entropy.

%% file: tsw.tex
\section{Thermal shallow water equations}
\label{sec: tsw}
In this section, we introduce the thermal shallow water equations and review their conservation properties.  
The thermal shallow water equations describe the evolution of fluid velocity $\boldsymbol{u}(\boldsymbol{x},t)$ and depth $\varphi(\boldsymbol{x},t)$ with respect to buoyancy transport.
There are two types of buoyancy transport \cite{Eldred2019,Ricardo2023,Ricardo2024-dg}. 
The first is material transport of buoyancy $b(\boldsymbol{x},t)=g\rho(\boldsymbol{x},t)/\bar{\rho}$ as a function of  density $\rho(\boldsymbol{x},t)$, vertically averaged density $\bar{\rho}$, and gravity $g$ \cite{Zeitlin2018}. 
The second is flux transport of density weighted buoyancy $B(\boldsymbol{x},t)=\varphi(\boldsymbol{x},t) b(\boldsymbol{x},t)$. 
Both support non-canonical Hamiltonian formulations  and reduce to the rotating shallow water equations when $b(\boldsymbol{x},t)=g$. 
We consider flux transport of $B(\boldsymbol{x},t)$. This yields the vector invariant form of the thermal shallow water equations for which the total energy, or Hamiltonian, is
\begin{align}
	\mathcal{H}(t) &= \int_\Omega \left( \frac{1}{2} \varphi(\boldsymbol{x},t) \boldsymbol{u}(\boldsymbol{x},t)\cdot \boldsymbol{u}(\boldsymbol{x},t) + \frac{1}{2}\varphi(\boldsymbol{x},t) B(\boldsymbol{x},t) \right).
\end{align}
 
\subsection{Continuous system}
Let $\Omega\subset \mathbb{R}^2$ be a two-dimensional  spatial domain with periodic boundaries. 
We use $^{\perp}$ to denote the 90 degree counterclockwise rotation of a two-dimensional vector in the plane. 
That is, if  $\boldsymbol{x}=(x_1,x_2)$, then $\boldsymbol{x}^\perp = \hat{\boldsymbol{k}}\times \boldsymbol{x} = (-x_2,x_1)$  where $\hat{\boldsymbol{k}}$ is the unit vector normal to plane, and $x_1$ and $x_2$ are the  horizontal and vertical coordinate respectively. Further, we write \reviewtwo{$\nabla(\cdot) = (\partial_{x_1}(\cdot),\partial_{x_2}(\cdot))$} in component form in order to introduce  \reviewtwo{$\nabla^\perp(\cdot)=(-\partial_{x_2}(\cdot),\partial_{x_1}(\cdot))$} as the skew gradient and  \reviewtwo{$\nabla^\perp\cdot(\cdot)=\hat{\boldsymbol{k}}\cdot \nabla \times(\cdot) $} as the two-dimensional analogue of curl, \reviewtwo{where $(\cdot)$ is an appropriate input function or vector.  }
Letting  $J\coloneqq (0,T), T>0$ represent the temporal domain, the vector invariant form of the thermal shallow water equations is
\begin{subequations}
	\begin{align}
		\frac{\partial \boldsymbol{u}(\boldsymbol{x},t)}{\partial t} + q(\boldsymbol{x},t) \boldsymbol{F}(\boldsymbol{x},t)^\perp + \nabla \Phi(\boldsymbol{x},t)  + b(\boldsymbol{x},t) \nabla \vartheta(\boldsymbol{x},t) &= 0, \qquad \text{in }\Omega \times  J ,
		\label{eq: u flux}\\
		\frac{\partial \varphi(\boldsymbol{x},t) }{\partial t} + \nabla \cdot \boldsymbol{F}(\boldsymbol{x},t) &= 0 ,
		\qquad \text{in }\Omega \times J , \label{eq: h flux}\\
		\frac{\partial B(\boldsymbol{x},t)}{\partial t} + \nabla \cdot \left( b(\boldsymbol{x},t) \boldsymbol{F}(\boldsymbol{x},t) \right) &= 0 , \qquad \text{in }\Omega \times J ,
		 \label{eq: B flux}
	\end{align}
\label{eq: tsw flux}
\end{subequations}
where   $q(\boldsymbol{x},t)$ is the potential vorticity. 
The mass flux $\boldsymbol{F}(\boldsymbol{x},t)$, Bernoulli potential $\Phi(\boldsymbol{x},t)$, and temperature $\vartheta(\boldsymbol{x},t)$, are the variational derivatives of $\mathcal{H}(t)$ with respect to $\boldsymbol{u}(\boldsymbol{x},t)$, $\varphi(\boldsymbol{x},t)$, $B(\boldsymbol{x},t)$, respectively. 
The \textit{prognostic} variables, $\boldsymbol{u}(\boldsymbol{x},t),\varphi(\boldsymbol{x},t),B(\boldsymbol{x},t)$, solve the time dependent problem \cref{eq: tsw flux}. The \textit{diagnostic} variables,  $\boldsymbol{F}(\boldsymbol{x},t),\Phi(\boldsymbol{x},t),\vartheta(\boldsymbol{x},t),q(\boldsymbol{x},t),b(\boldsymbol{x},t)$, solve the algebraic constraints
\begin{align}
	\boldsymbol{F}(\boldsymbol{x},t) &= \varphi(\boldsymbol{x},t) \boldsymbol{u}(\boldsymbol{x},t) ,& 
	\Phi(\boldsymbol{x},t) &= \frac{1}{2}\boldsymbol{u}(\boldsymbol{x},t)\cdot \boldsymbol{u}(\boldsymbol{x},t) + \frac{1}{2}B(\boldsymbol{x},t), & 
	\vartheta(\boldsymbol{x},t)  &= \frac{1}{2}\varphi(\boldsymbol{x},t), \nonumber \\
	q(\boldsymbol{x},t) &= \frac{\nabla^\perp \cdot \boldsymbol{u}(\boldsymbol{x},t) + f }{\varphi(\boldsymbol{x},t)} ,& 
	b(\boldsymbol{x},t) &= \frac{B(\boldsymbol{x},t)}{\varphi(\boldsymbol{x},t)}, \label{eq: constraints}
\end{align}
where $f$ is the Coriolis parameter. 

\subsection{System invariants}
\label{sec: system invariants}

Using a skew symmetric operator to express \cref{eq: tsw flux} in terms of $\mathcal{H}(t)$ means energy is conserved \cite{Eldred2019,Ricardo2023,Ricardo2024-dg,Wimmer2020,BauerCotter2018}.
Skew symmetric operators, which represent non-canonical Poisson brackets for compressible fluids, have a nullspace that contains the variational derivatives of additional system invariants, known as Casimirs \cite{Shepherd1990,Salmon1998}. 
These include total mass $\mathcal{M}(t) = \int_\Omega \varphi(\boldsymbol{x},t)$, buoyancy $\mathcal{B}(t)= \int_\Omega B(\boldsymbol{x},t)$,  vorticity $\mathcal{V}(t) = \int_\Omega \varphi(\boldsymbol{x},t) q(\boldsymbol{x},t)$ and higher order moments of buoyancy such as entropy
\begin{align}
	\mathcal{S}(t) &=  \int_\Omega \frac{1}{2} b(\boldsymbol{x},t)^2 \varphi(\boldsymbol{x},t) .
	\label{eq: entropy}
\end{align}

To  motivate the discrete approximation  proposed in \cref{sec: discretisation,sec: temp discretisation} below, we discuss entropy conservation at the continuous level. Hereafter, we drop the dependence on space and time. 
\reviewonee{
Assuming $b,\varphi,B \in H^1(J)$, differentiating  \cref{eq: entropy} with respect to time yields the total time derivative of $\mathcal{S}$ as
\begin{subequations}
	\begin{align}
		\frac{\mathrm{d}\mathcal{S}}{\mathrm{d}t} 
		&= 
		\int_{\Omega} b\varphi \frac{\partial b}{\partial t} 
		+ \int_{\Omega} \frac{1}{2}b^2 \frac{\partial \varphi}{\partial t}, \label{eq: dsdt 1} \\
		&=  \int_{\Omega} B \frac{\partial }{\partial t} \left( B\varphi^{-1} \right) 
		+ \int_{\Omega} \frac{1}{2} b^2 \frac{\partial \varphi }{\partial t}, \label{eq: dsdt 2 a}  \\
		&= \int_{\Omega} B\varphi^{-1} \frac{\partial B }{\partial t} 
		- \int_{\Omega} B^2 \varphi^{-2} \frac{\partial \varphi}{\partial t}
		+ \int_{\Omega} \frac{1}{2} b^2 \frac{\partial \varphi }{\partial t} , \label{eq: dsdt 2} \\
		&= \int_{\Omega} b \frac{\partial B }{\partial t} 
		- \int_{\Omega} \frac{1}{2}b^2 \frac{\partial \varphi}{\partial t} . \label{eq: dsdt pre}
	\end{align}
\end{subequations} 
To obtain  \cref{eq: dsdt 2 a}, we substitute the relation $b=B \varphi^{-1}$ into the first term of \cref{eq: dsdt 1}. 
Expanding the temporal derivatives in the first term of \cref{eq: dsdt 2 a}  gives \cref{eq: dsdt 2}. Further substituting $b=B \varphi^{-1}$ into the first and second terms of \cref{eq: dsdt 2} and simplifying yields  \cref{eq: dsdt pre}. 
An equivalent expression for \cref{eq: dsdt pre} is  
\begin{align}
	\frac{\mathrm{d}\mathcal{S}}{\mathrm{d}t} =  \int_{\Omega}\frac{\delta \mathcal{S}}{\delta B} \frac{\partial B }{\partial t} 
	+  \int_{\Omega} \frac{\delta \mathcal{S}}{\delta \varphi} \frac{\partial \varphi}{\partial t}
	,  \label{eq: dsdt}
\end{align} 
where  $\delta \mathcal{S}/\delta B=b$ and $\delta \mathcal{S}/\delta \varphi=-b^2/2$  are the variational derivatives of entropy. 
}

Next, substituting  \cref{eq: h flux,eq: B flux} into \cref{eq: dsdt pre} yields
\reviewonee{
\begin{subequations}
\begin{align}
	\frac{\mathrm{d} \mathcal{S}}{\mathrm{d}t}
	&=   \int_\Omega  \frac{1}{2} b^2 \nabla \cdot \boldsymbol{F}  - \int_\Omega b \nabla \cdot (b \boldsymbol{F}),  \label{eq: dsdt space 1}
	\\
	&=   \int_\Omega  \frac{1}{2} b^2 \nabla \cdot \boldsymbol{F}  
	- \int_\Omega b \left( b \nabla \cdot \boldsymbol{F} + \boldsymbol{F}\cdot \nabla b  \right),  \label{eq: dsdt space 2}   \\
	&=   -\int_\Omega \frac{1}{2} b^2 \nabla \cdot \boldsymbol{F}  
	- \int_\Omega b\boldsymbol{F}\cdot \nabla b  ,  \label{eq: dsdt space 3}
\end{align}
\end{subequations}
where \cref{eq: dsdt space 2} results from expanding the spatial derivatives in the second term of \cref{eq: dsdt space 1}, and \cref{eq: dsdt space 3} arises from grouping terms. 
Now we assume  $\nabla (b^2) \in H^1(\Omega)$, $\boldsymbol{F}\in H(\mathrm{div},\Omega)$ in order to apply integration by parts to the first term in  \cref{eq: dsdt space 3}.
}
Expanding  spatial derivatives in the resulting term  gives 
\begin{align}
	\frac{\mathrm{d} \mathcal{S}}{\mathrm{d}t}
	&=   \int_\Omega \frac{1}{2}\nabla (b^2 ) \cdot \boldsymbol{F}  
	- \int_\Omega b\boldsymbol{F}\cdot \nabla b     
	=   \int_\Omega b\boldsymbol{F}\cdot \nabla b  
	- \int_\Omega b\boldsymbol{F}\cdot \nabla b     
	= 0 .\label{eq: dsdt space}
\end{align}

The above analysis is used to determine the regularity requirements for discrete entropy conservation. 
For \cref{eq: dsdt} to hold discretely in time, we consider a temporally continuous approximation of $b$ and the prognostic variables. 
Thus, \reviewtwo{entropy is a continuous cubic polynomial} in time, which is not conserved in general at the discrete level by the chosen time integrator \cite{Cohen2011}.   
The  spatial approximation of prognostic variables  is determined by a discrete de Rham complex \cite{Cotter2023}, where $\nabla \cdot \boldsymbol{u}, \varphi,B$ are spatially discontinuous.
For the vector invariant form of the thermal shallow water equations,  $b$ is typically also spatially discontinuous \cite{Ricardo2023,Eldred2019}. 
This choice means entropy conservation in \cref{eq: dsdt space} does not hold at the semi-discrete level \cite{Ricardo2023}.  
While the material form of the thermal shallow water equations facilitates spatially continuous approximations of $b$,  the associated variational derivative of the Hamiltonian is not collocated with the pressure gradient, and energy is not readily conserved  \cite{Eldred2019}.

To allow for spatially discontinuous approximations of $b$,  we define a numerical approximation of \cref{eq: tsw flux} such that integration by parts is not required  to prove semi-discrete entropy conservation. 
Using the \reviewtwo{product rule} to reformulate the buoyancy equation \cref{eq: B flux} and reciprocating modifications in  the momentum equation \cref{eq: u flux}, we restate \cref{eq: tsw flux} as \cite{Ricardo2023}
\begin{subequations}\label{eq: tsw expanded}
	\begin{align}
		\frac{\partial \boldsymbol{u}}{\partial t} + q \boldsymbol{F}^\perp + \nabla \Phi + \frac{1}{2}b \nabla \vartheta 
		+ \frac{1}{2}\nabla(b\vartheta) - \frac{1}{2} \vartheta \nabla b &=0,  \label{eq: tsw expanded u} \\
		\frac{\partial \varphi}{\partial t} + \nabla \cdot \boldsymbol{F}  &= 0,  \label{eq: tsw expanded h} \\
		\frac{\partial B}{\partial t} + \frac{1}{2} \nabla \cdot (b\boldsymbol{F}) + \frac{1}{2} b\nabla \cdot \boldsymbol{F} + \frac{1}{2} \boldsymbol{F}\cdot \nabla b &=0. \label{eq: tsw expanded B}
	\end{align} 
\end{subequations}

Skew-symmetry is maintained in \cref{eq: tsw expanded} so that energy is  conserved, assuming periodic spatial boundary conditions. 
Substituting \cref{eq: tsw expanded h,eq: tsw expanded B} into \cref{eq: dsdt pre}, yields entropy conservation at the continuous level as in \cref{eq: dsdt space}.
At the discrete level, we apply integration by parts to $\nabla \cdot (b\boldsymbol{F})$ in \cref{eq: tsw expanded B} and account for jumps across interior element boundaries such that the resulting term readily cancels with the jumps related to the $\boldsymbol{F}\cdot \nabla b$ term in \cref{eq: tsw expanded B}. 
The $b\nabla \cdot \boldsymbol{F}$ term in \cref{eq: tsw expanded B} is div-conforming at the discrete level, and does not contribute to interior element boundaries. 
To  ensure discrete energy and entropy conservation, we design  skew symmetric jump terms by exploiting the regularity  of $b$ at the continuous level.

%% file: fem.tex
\section{Semi-discrete formulation}
\label{sec: discretisation}

We now present a compatible  finite element formulation of the thermal shallow water equations \cref{eq: tsw expanded}, and prove semi-discrete energy and entropy conservation. 
\reviewone{ 
Skew symmetric upwinded numerical fluxes  are  proposed to smooth spurious oscillations that can occur as simulations progress  to mature turbulent states. 
We subsequently prove such fluxes conserve energy and monotonically damp entropy. }

\subsection{Notation}

Let $\mathcal{T}_{h}^x$ be a conforming partition of the spatial domain $\Omega$ into quadrilaterals or triangles.
Let $(\cdot,\cdot)$ denote the inner product over elements in $\mathcal{T}_h^x$  and $\langle\cdot,\cdot \rangle$ denote the inner product over interior edges.
For neighbouring spatial elements $K^\pm \in \mathcal{T}_h^x$,  which share a common edge $e$, the jump, $\llbracket \cdot \rrbracket$, and average, $\left\{\cdot\right\}$, of a function on $e$ is  \cite{Brezzi2004}
\begin{align*}
	\llbracket x \rrbracket &= x^+ \boldsymbol{n}^+ + x^- \boldsymbol{n}^- ,& 
	\left\{x \right\} &= \frac{1}{2}\left( x^+ + x^- \right) \text{ on } e\in \mathcal{E}_0, \\
	\reviewonee{
	\llbracket \boldsymbol{y} \rrbracket} &= \reviewonee{\boldsymbol{y} ^+ \cdot \boldsymbol{n}^+ + \boldsymbol{y}^- \cdot \boldsymbol{n}^- },
	& 
	\left\{\boldsymbol{y} \right\} &= \frac{1}{2}\left( \boldsymbol{y}^+ + \boldsymbol{y}^- \right) \text{ on } e\in \mathcal{E}_0 ,
\end{align*}
where $\mathcal{E}_0$ is the set of interior edges, 
$x$ is a scalar valued function, $\boldsymbol{y}$ is a vector valued function,  $\cdot^\pm$ is the restriction to $K^\pm$, and $\boldsymbol{n}^+ = - \boldsymbol{n}^-$ is the outward pointing spatial normal vectors of $K^\pm$ on $e$. 

\subsection{Finite element spaces}

Spatial exterior derivatives,  $\nabla^\perp$ and $\nabla \cdot$,  map  compatible finite element spaces,
$\mathbb{V}_0 \subset H^1(\Omega)$, $\mathbb{V}_1 \subset H(\mathrm{div},\Omega)$, $\mathbb{V}_2 \subset L^2(\Omega)$,   via the two-dimensional discrete de Rham complex \cite{Cotter2023,Arnold2018}
\begin{align*}
	\begin{CD}
		\mathbb{V}_{0} \subset H^1(\Omega)
		@>   \nabla^{\perp}  >>
		\mathbb{V}_{1} \subset H(\mathrm{div},\Omega)
		@>  \nabla\cdot >>
		\mathbb{V}_{2} \subset L^2(\Omega).
	\end{CD}  
\end{align*}
We consider quadrilateral meshes where $\mathbb{V}_0 = \mathcal{P}_{p+1}(\mathcal{T}_{h}^x)$ is the scalar continuous Lagrangian finite element space of piecewise polynomials of order $p+1$, $\mathbb{V}_1=\mathcal{RT}_p(\mathcal{T}_{h}^x)$ is the polynomial Raviart-Thomas space of  order $p$ \cite{RT1977}, and $\mathbb{V}_2=\mathcal{P}_{p}^-(\mathcal{T}_{h}^x)$ is the space of piecewise discontinuous polynomials of order $p$. 
The  discussion hereafter applies to any other choice of compatible finite element spaces for quadrilateral and triangular meshes \cite{Arnold2018}.

\subsection{Finite element approximation}
\label{sec: discrete system}

The semi-discrete finite element formulation  of the thermal shallow water equations in \cref{eq: tsw expanded} is:
find $\boldsymbol{u}_h \in \mathbb{V}_{1}$, $\varphi_h, B_h \in \mathbb{V}_{2}$,  such that
\begin{subequations}
	\begin{align}
		\left( \frac{ \partial \boldsymbol{u}_h}{\partial t} , \psi_{\boldsymbol{u}_h} \right) 
		&+ \left(  q_h ,\boldsymbol{F}_h^\perp \cdot \psi_{\boldsymbol{u}_h} \right) 
		- 	\left( \nabla  \cdot \psi_{\boldsymbol{u}_h}, \Phi_h \right)  \nonumber \\
		&-g ( \psi_{\boldsymbol{u}_h},b_h,\widetilde{b_h},\vartheta_h  )  
		- s(\psi_{\boldsymbol{u}_h},b_h,\vartheta_h)
		=0, \label{eq: fem u}
		&\forall \psi_{\boldsymbol{u}_h} \in \mathbb{V}_{1} ,   \\
		\left( \frac{ \partial \varphi_h}{\partial t}, \phi_{\varphi_h}\right) & + \left( \nabla \cdot \boldsymbol{F}_h , \phi_{\varphi_h} \right) = 0, \label{eq: fem h}   
		& 	\forall \phi_{\varphi_h} \in \mathbb{V}_{2} , \\
		\left( \frac{ \partial  B_h}{\partial t},  \phi_{B_h} \right)
		&+	g ( \boldsymbol{F}_h,b_h,\widetilde{b_h},\phi_{B_h}  )   +s(\boldsymbol{F}_h,b_h,\phi_{B_h})
		=0,     \label{eq: fem B}
		&\forall \phi_{B_h} \in \mathbb{V}_{2} ,
	\end{align} 
	\label{eq: fem}
\end{subequations}
together with periodic spatial boundary conditions. 
The forms $g(\cdot,b_h,\widetilde{b_h},\cdot)$ and $s(\cdot,b_h,\cdot) = s_\mathrm{c}(\cdot,b_h,\cdot) + s_{\mathrm{up}}(\cdot,b_h,\cdot)$ are the discretisation of the advection and stabilisation terms, which consists of a centred and upwinded flux, as follows:
\begingroup 
	\allowdisplaybreaks
\begin{subequations}
	\begin{align}
		g ( \boldsymbol{w}_h,b_h,\widetilde{b_h},\phi_h  ) 
		&= -   \frac{1}{2} \left( b_h, \boldsymbol{w}_h \cdot \nabla_h  \phi_h \right)
		+   \frac{1}{2} \left(  \widetilde{b_h} \phi_h, \nabla \cdot \boldsymbol{w}_h  \right)
		+   \frac{1}{2} \left( \phi_h,   \nabla_h b_h \cdot\boldsymbol{w}_h \right) ,\\
		s_{\mathrm{c}}(\boldsymbol{w}_h,b_h,\phi_h)
		&=   \frac{1}{2} \left\langle  \left\{\boldsymbol{w}_h b_h \right\} ,  \llbracket \phi_h \rrbracket  \right\rangle 
		-  \frac{1}{2}
		\left\langle 
		\left\{\boldsymbol{w}_h \phi_h \right\},  \llbracket b_h \rrbracket  \right\rangle , \\
		s_{\mathrm{up}}(\boldsymbol{w}_h,b_h,\phi_h)
		&=  \frac{1}{2} 
		\left\langle \alpha(\boldsymbol{w}_h)  \llbracket \phi_h  \rrbracket,  \llbracket b_h \rrbracket \right\rangle ,\label{eq: forms upwind}
	\end{align}
	\label{eq: forms}
\end{subequations}
\endgroup
for all $\boldsymbol{w}_h \in \mathbb{V}_{1}$, $\phi_h \in \mathbb{V}_{2}$, where
$\widetilde{b_h}$ is a specific formulation of the buoyancy introduced in \cref{eq: b tilde time} below,
and $\alpha(\boldsymbol{w}_h) \coloneqq  |\boldsymbol{w}_h\cdot \boldsymbol{n}^+ |/2$ is the upwinding parameter \cite{Brezzi2004}. 
Centred fluxes correspond to $\alpha = 0$ and  upwinded fluxes correspond to $\alpha > 0$.
The upwinded fluxes in \cref{eq: forms upwind} align with other studies that consider hyperbolic systems involving advection \cite{Brezzi2004,Eldred2019}, and maintain skew symmetry in \cref{eq: fem}. 
Upwinding is not required in the continuity equation since  the corresponding linearised equation is balanced with respect to left and right solutions \cite{Vallis2006}, and $\boldsymbol{F}_h \in \mathbb{V}_1$ is continuous across element boundaries.

The semi-discrete approximation of the diagnostic variables in \cref{eq: constraints} is: find $\boldsymbol{F}_h \in \mathbb{V}_{1}$, $\Phi_h,\vartheta_h, b_h \in \mathbb{V}_{2}$, $q_h \in \mathbb{V}_{0}$,  such that
\begin{subequations}
	\begin{align}
		\left(\boldsymbol{F}_h ,  \psi_{\boldsymbol{F}_h}\right)  &=\left(  \varphi_h \boldsymbol{u}_h , \psi_{\boldsymbol{F}_h} \right) ,\label{eq: fem F} 	
		&\forall \psi_{\boldsymbol{F}_h} \in \mathbb{V}_{1} , \\
		\left( \Phi_h , \phi_{\Phi_h}\right)  &=	    \left(\frac{1}{2} \boldsymbol{u}_h\cdot \boldsymbol{u}_h +  \reviewone{ \frac{1}{2}} B_h , \phi_{\Phi_h}\right) , \label{eq: fem phi}
		&\forall \phi_{\Phi_h} \in \mathbb{V}_{2} ,\\
		\left( \vartheta_h ,\phi_{\vartheta_h} \right)  &=    \left( \frac{1}{2}\varphi_h ,\phi_{\vartheta_h}\right)  , \label{eq: fem T}
		&\forall \phi_{\vartheta_h} \in \mathbb{V}_{2} , \\
		\left(	b_h  \varphi_h, \phi_{b_h} \right)&= \left(  B_h , \phi_{b_h} \right), 
		&\forall \phi_{b_h} \in \mathbb{V}_2,
		\label{eq: fem b semi}  \\
		\left( q_h \varphi_h , \xi_{q_h} \right)  &= 	- \left( \nabla^\perp \xi_{q_h}, \boldsymbol{u}_h \right)   + \left( f, \xi_{q_h} \right) , \label{eq: fem q}
		&\forall \xi_{q_h} \in \mathbb{V}_{0} ,
	\end{align}
	\label{eq: fem diagnostic}
\end{subequations}
where \cref{eq: fem T} holds pointwise  and is not required in practice. 
The final diagnostic is:
find $\widetilde{b_h} \in  \mathbb{V}_{2}$,  such that
\begin{align}
	\left(   \widetilde{b_h} b_h ,\widetilde{\phi}_h\right)  &= \left(b_hb_h, \widetilde{\phi}_h\right) ,  \label{eq: b tilde semi}
	& \forall \widetilde{\phi}_h \in \mathbb{V}_{2},
\end{align}
where the projection of the quadratic term $b_hb_h$ into $\mathbb{V}_2$ is accounted for since $\widetilde{\phi}_h \in \mathbb{V}_{2}$.
While entropy can be conserved at the semi-discrete level by replacing  $\widetilde{b_h}$ with $b_h$ \cite{Ricardo2023,Ricardo2024-dg}, 
we show in \cref{sec: temp conservation}  that the additional diagnostic  \cref{eq: b tilde semi} 
ensures the forcing terms exactly conserve entropy in time. Thus, the only source of temporal conservation error is the approximation of  time derivatives.

We now prove the semi-discrete formulation is a consistent finite element approximation of the thermal shallow water equations that supports compatible advection of buoyancy.
\begin{proposition}
	\label{prop: consistency}
	Solutions of the continuous system  are consistent with the semi-discrete formulation. 
	\begin{proof}
		To show smooth solutions of the continuous system satisfy the semi-discrete system, discrete quantities  in \cref{eq: fem} are replaced by their continuous analogue. Equation  
		\cref{eq: fem} contains terms involving the time derivative of the prognostic variables, for which the consistency is trivial, and all other terms relate to the diagnostic variables.  
		We now elaborate on the consistency of the forms $g(\cdot,b_h,\widetilde{b_h},\cdot) $ and $s(\cdot,b_h,\cdot)$. 
		
		First we observe that  $\llbracket b\rrbracket = 0$, $\widetilde{b} = b$, and $\llbracket \vartheta \rrbracket = 0$  for smooth solutions at the continuous level. 
		For the momentum equation \cref{eq: fem u}, the terms in question reduce to   
		\begin{align*}
			-g(\psi_{\boldsymbol{u}},b,b,\vartheta )- s(\psi_{\boldsymbol{u}},b,\vartheta )
			&= -g(\psi_{\boldsymbol{u}},b,b,\vartheta ) 
			=  \left( b,  \nabla  \vartheta \cdot \psi_{\boldsymbol{u}} \right) ,
		\end{align*}
		after applying integration by parts and the chain rule. 
		Consistency readily follows noting that $\nabla \cdot \psi_{\boldsymbol{u}} \Phi$ is div-conforming. 
		\reviewone{A similar procedure applies to the density weighted buoyancy equation \cref{eq: fem B}.    }
	\end{proof}
\end{proposition}

\begin{proposition}
	The semi-discrete formulation supports compatible advection of buoyancy.  
	\begin{proof}
		As stated by \citet{Eldred2019}, proving compatible advection of buoyancy is equivalent to showing the density weighted buoyancy equation \cref{eq: fem B} reduces to the continuity equation \cref{eq: fem h} when $b_h = 1$. 
		
		Setting $b_h=1$ in \cref{eq: fem b semi,eq: b tilde semi} yields $\varphi_h=B_h$ pointwise and $\widetilde{b_h}=1$ respectively. 
		Next taking $b_h=\widetilde{b_h}=1$ in \cref{eq: fem B}, and expanding $g ( \boldsymbol{F}_h,1,1,\phi_{B_h}  ) $  and $s(\boldsymbol{F}_h,1,\phi_{B_h})$ gives 		
		\begin{align} 
				\left( \frac{ \partial  B_h}{\partial t},  \phi_{B_h} \right)
				&
				+   \frac{1}{2} \left(   \phi_h, \nabla \cdot \boldsymbol{F}_h  \right)
				-   \frac{1}{2} \left( 1, \boldsymbol{F}_h \cdot \nabla_h  \phi_h \right)
				+  \frac{1}{2} \left\langle  \left\{\boldsymbol{F}_h   \right\} ,  \llbracket \phi_h \rrbracket  \right\rangle 
				= 0,
				&\forall \phi_{B_h} \in \mathbb{V}_{2} ,
		\end{align}
		since $\llbracket 1\rrbracket =\nabla_h 1 = 0 $ and $\left\{ 1\right\} = 1$. Applying integration by parts   yields \cref{eq: fem h} since $\varphi_h = B_h$. 
	\end{proof} 
\end{proposition}

\subsection{Semi-discrete conservation}
\label{sec: conservation}
We now prove the main result of this study:  the semi-discrete system \cref{eq: fem}  conserves discrete entropy when $\alpha = 0$, monotonically damps entropy when $\alpha > 0$, and conserves energy always.


\begin{proposition}
	\label{prop: energy conservation}
	The semi-discrete system  conserves energy.
	\begin{proof}
		Discrete energy at any time $t$  is
		\begin{align}
			\mathcal{H}_h(t)   &=   \int_\Omega \left( \frac{1}{2} \varphi_h \boldsymbol{u}_h \cdot \boldsymbol{u}_h + \frac{1}{2}\varphi_h B_h  \right), \qquad \forall t \in J. \label{eq: discrete energy t}   
		\end{align}
		Differentiating $\mathcal{H}_h(t) $ with respect to time  gives
		\begin{equation}
			\begin{aligned}    
				\frac{\partial \mathcal{H}}{\partial t} &=    \left( \frac{
				\delta \mathcal{H}_h}{\delta \boldsymbol{u}_h} , 
				\frac{\partial \boldsymbol{u}_h}{\partial t} \right) 
				+ 
				\left( \frac{ \delta \mathcal{H}_h}{\delta \varphi_h } , \frac{ \partial \varphi_h}{\partial t}   \right) 
				+ 
				\left( \frac{ \delta \mathcal{H}_h}{\delta B_h }, \frac{\partial B_h}{\partial t} \right)  ,
				\label{eq: dHdt}
			\end{aligned} 
		\end{equation}	
	where  
		\begin{subequations}
			\begin{align}
			\frac{ \delta\mathcal{H}_h }{\delta \boldsymbol{u}_h}\in \mathbb{V}_1: &&
			 \left(   \frac{ \delta\mathcal{H}_h }{\delta \boldsymbol{u}_h} ,   \phi_{\boldsymbol{u}_h}\right) &= \left(   \varphi_h \boldsymbol{u}_h ,  \phi_{\boldsymbol{u}_h}  \right)
			 = 
			 \left(   \boldsymbol{F}_h ,  \phi_{\boldsymbol{u}_h}  \right) , & \forall \phi_{\boldsymbol{u}_h} \in \mathbb{V}_{1},   \\
			\frac{ \delta \mathcal{H}_h}{\delta \varphi_h} \in \mathbb{V}_2:&&
				\left(  \frac{ \delta \mathcal{H}_h}{\delta \varphi_h}, \phi_{\varphi_h} \right) &=
				\left(  \frac{1}{2} \boldsymbol{u}_h \cdot \boldsymbol{u}_h + \reviewone{\frac{1}{2}} B_h ,\phi_{\varphi_h}\right)
				=\left( \Phi_h ,\phi_{\varphi_h}\right)      , & \forall \phi_{\varphi_h} \in \mathbb{V}_{2}, \\
			\frac{ \delta \mathcal{H}_h}{\delta B_h} \in \mathbb{V}_2: &&
				\left(   \frac{ \delta \mathcal{H}_h}{\delta B_h}, \phi_{B_h}\right) &=
				\left(\frac{1}{2}\varphi_h , \phi_{B_h}\right)
				= \left(\vartheta_h , \phi_{B_h}\right)   , & \forall \phi_{B_h} \in \mathbb{V}_{2} , 		 
			\end{align}	
			\label{eq: var der H st}
		\end{subequations}   
		are the functional derivatives of $\mathcal{H}_h(t)$ that  hold pointwise. 		
		Substituting  \cref{eq: var der H st} into  \cref{eq: dHdt}
		and evaluating the resulting expression using  $\psi_{\boldsymbol{u}_h} = \boldsymbol{F}_h$, $\phi_{\varphi_h} = \Phi_h$, $\phi_{B_h} = \vartheta_h$
		in  \cref{eq: fem} gives ${\partial\mathcal{H}_h}/{\partial t} =0$. 	
	\end{proof}
\end{proposition}


\begin{proposition}
	\label{prop: entropy conservation semi}
	The semi-discrete  system conserves entropy when $\alpha = 0$, and monotonically damps entropy otherwise for $\alpha > 0$. 
	\begin{proof} 
		Discrete entropy at any time  $t$  is
		\begin{align}
			\mathcal{S}_h (t)   &=  \int_{\Omega}  \frac{1}{2} b_h b_h \varphi_h , 
			\qquad \forall t \in J.
			\label{eq: discrete entropy semi}
		\end{align}
		Under the assumption of \reviewtwo{continuity in time}, \cref{eq: dsdt} gives the time derivative of $\mathcal{S}_h$ as
		\begin{align}
		  \frac{ \partial \mathcal{S}_h}{\partial t}  &= 
			\left( \frac{  \delta \mathcal{S}_h}{\delta \varphi_h},\frac{ \partial \varphi_h}{\partial t} \right)
			+  \left( \frac{ \delta   \mathcal{S}_h}{\delta {B_h}}, \frac{\partial B_h}{\partial t} \right).
			\label{eq: dSdt semi}
		\end{align}
		
		In contrast to energy, entropy involves both prognostic and diagnostic variables. Evaluating the discrete  functional derivatives requires several steps as follows.  
		First differentiating \cref{eq: discrete entropy semi} with respect to $\varphi_h$ and $b_h$ gives 
		\begin{subequations}
			\begin{align}
				\left(  \frac{\partial \mathcal{S}_h}{\partial \varphi_h}, \phi_{\varphi_h} \right)
				&= \left( \frac{1}{2}  b_h b_h ,\phi_{\varphi_h}\right) ,  
				& \forall \phi_{\varphi_h}  \in \mathbb{V}_{2} ,  \label{eq: dSdh h semi}\\
				\left(  \frac{\partial \mathcal{S}_h}{\partial b_h}, \phi_{b_h} \right) &= \left( b_h \varphi_h, \phi_{b_h} \right) ,
				& \forall \phi_{b_h}  \in \mathbb{V}_{2}.
				\label{eq: dSdb semi}
			\end{align}
			\label{eq: dS semi}
		\end{subequations} 
		Further differentiating \cref{eq: fem b semi} with respect to $\varphi_h$ and $B_h$ yields
		\begin{subequations}
			\begin{align}
				\left(  \frac{\partial b_h}{\partial \varphi_h}  	\varphi_h , \phi_{b_h} \right) &=  \left( - b_h\phi_{\varphi_h} ,\phi_{b_h}\right) , 
				& \forall \phi_{b_h},\phi_{\varphi_h} \in \mathbb{V}_2 , 
				\label{eq: dbdh semi}
				\\
				\left( \frac{\partial b_h}{\partial B_h}  	\varphi_h ,\phi_{b_h}\right)  &= \left(  \phi_{B_h},  \phi_{b_h} \right) , 
				&\forall \phi_{b_h},\phi_{B_h}  \in \mathbb{V}_2 .\label{eq: dbdB semi}
			\end{align}
			\label{eq: db}
		\end{subequations} 
	\reviewtwo{To obtain an expression for $\delta \mathcal{S}_h/\delta \varphi_h$, we set $\phi_{b_h} = {\partial b_h}/{\partial \varphi_h}$ in \cref{eq: dSdb semi}, $\phi_{\varphi_h}=b_h$  in \cref{eq: dbdh semi},  and combine the result with \cref{eq: dSdh h semi}.	
	Deriving an expression for $\delta \mathcal{S}_h/\delta B_h$ requires taking $\phi_{b_h} = {\partial b_h}/{\partial B_h}$ in \cref{eq: dSdb semi} and $\phi_{b_h}=b_h$  in \cref{eq: dbdB semi}.}	Thus, the variational derivatives of $\mathcal{S}_h(t)$  are:
		\begin{subequations}
			\begin{align}
				\text{Find }  \frac{\delta \mathcal{S}_h}{\delta \varphi_h}\in \mathbb{V}_{2} : 
				\left(  \frac{\delta \mathcal{S}_h}{\delta \varphi_h}, \phi_{\varphi_h} \right)
				&= 	 \left( \frac{\partial \mathcal{S}_h}{\partial b_h} ,\frac{\partial b_h }{\partial \varphi_h } \right)
				+ \left( \frac{\partial \mathcal{S}_h}{\partial \varphi_h}, \phi_{\varphi_h} \right)
				=  \left( - \frac{1}{2} b_h b_h , \phi_{\varphi_h}  \right) , \label{eq: dSdh semi}
				~\forall \phi_{\varphi_h} \in \mathbb{V}_{2}, 
				\\
				\text{Find } \frac{\delta \mathcal{S}_h}{\delta B_h}\in \mathbb{V}_{2} : 
				\left( \frac{\delta \mathcal{S}_h}{\delta B_h}, \phi_{B_h} \right) &= 	\left( \frac{\partial \mathcal{S}_h}{\partial 	b_h} ,\frac{\partial b_h }{\partial B_h }\right) 
				= \left( b_h , \phi_{B_h}\right)  , 
				~~~~\forall \phi_{B_h} \in \mathbb{V}_{2} . \label{eq: dSdB semi}
			\end{align}  
			\label{eq: var dervs S semi}
		\end{subequations}

		To evaluate  \cref{eq: dSdt semi}, 
		set  $\phi_{\varphi_h} = -\Pi_{\mathbb{V}_2}(b_h b_h)/2 $ in \cref{eq: fem h} and  $\phi_{B_h} = b_h$ in \cref{eq: fem B},
		where $\Pi_{\mathbb{V}_2}$ is the projection into $\mathbb{V}_2$.
		Expanding $g(\boldsymbol{F}_h,b_h,\widetilde{b_h},b_h)$ and $s(\boldsymbol{F}_h,b_h,b_h)$ in the resulting expression shows integration by parts is not required to cancel terms. 
		Thus, 
			\begin{align}
				 \frac{ \partial \mathcal{S}_h}{\partial t} 
				&=  \frac{1}{2} \left( b_h b_h,  \nabla \cdot \boldsymbol{F}_h \right)    
				- \frac{1}{2} \left(  \widetilde{b_h}	{b_h},  \nabla \cdot \boldsymbol{F}_h \right) 
				-s_{\mathrm{up}}(\boldsymbol{F}_h,{b_h},{b_h}), 
				\label{eq: dsdt semi 2}
			\end{align} 
		where the projection into $\mathbb{V}_2$ cancels since $\nabla \cdot \boldsymbol{F}_h \in \mathbb{V}_2$. 
		Using $\widetilde{\phi}_h =\nabla \cdot \boldsymbol{F}_h$ in \cref{eq: b tilde semi} 
		yields $ { \partial \mathcal{S}_h}/{\partial t}   =0$ when $\alpha = 0$,  and $  { \partial \mathcal{S}_h}/{\partial t}   < 0 $ when $\alpha > 0 $  since  $s_{\mathrm{up}}(\boldsymbol{F}_h,{b_h},{b_h})>0$ for  $\alpha > 0$.
	\end{proof}
\end{proposition}

In \cref{prop: entropy conservation semi,prop: energy conservation}, we prove the centred fluxes in the semi-discrete approximation are energy--entropy conserving, and the upwinded fluxes are energy-conserving--entropy-damping. 
The fact that integration by parts is not required to cancel terms in \cref{eq: dsdt semi 2} is a direct consequence of the reformulation of the continuous system \cref{eq: tsw expanded} to contain  the expanded chain rule \cite{Ricardo2023}. 
This approach was previously reported by 
\citet{Ricardo2023,Ricardo2024-dg}, who also used a similar reformulation of continuous system to motivate semi-discrete entropy conservation. 
A key contribution here is that the condition on $\widetilde{b_h}$ in \cref{eq: b tilde semi}  provides the necessary cancellation for entropy conservation in \cref{eq: dsdt semi 2} when $\alpha = 0$, and the upwinded fluxes are constructed to  monotonically damp entropy when $\alpha > 0 $. 
Thus, we achieve stabilised entropy conservation, and  unify other entropy conserving studies that consider mixed finite elements without stabilisation \cite{Ricardo2023} and the discontinuous Galerkin method \cite{Ricardo2024-dg}.  
Moreover, discrete energy conservation results from the skew symmetry of \cref{eq: fem} that is preserved by the occurrence of $\widetilde{b_h}$ in both the momentum and buoyancy equations, and holds for all  $\alpha$  since the stabilisation terms are skew symmetric.

Discrete mass $\mathcal{M}_h(t) = \int_\Omega \varphi_h$ is conserved since setting $\phi_{\varphi_h}=1$ in \cref{eq: fem h} results in a flux form conservation law. 
Taking $\phi_{B_h}=1$ in \cref{eq: fem B} and applying integration by parts reveals the discrete total buoyancy, $\mathcal{B}_h= \int_\Omega B_h$, is only conserved when $\widetilde{b_h}=b_h$ pointwise.  
Similar to other finite element approximations of the thermal shallow water equations \cite{Eldred2019}, discrete vorticity, $\mathcal{V}_h = \int_\Omega \varphi_h q_h$, is conserved when $b_h$ is a constant.
Thus, the semi-discrete formulation satisfies the criteria of a compatible  spatial discretisation for the thermal shallow water equations \cite{Eldred2019,Ricardo2024-dg}.


\section{Fully discrete formulation}
\label{sec: temp discretisation}

In this section, we integrate the semi-discrete system \cref{eq: fem} in time to develop a fully discrete formulation of the thermal shallow water equations. Temporal conservation is also analysed.  


\subsection{Poisson integrator}
\reviewtwo{Since the semi-discrete system \cref{eq: fem} originates from a skew symmetric formulation of the thermal shallow water equations, it is a   Poisson system of the form}
\begin{align}
	\frac{\partial \boldsymbol{z}_h}{\partial t} = \mathbb{A}(\boldsymbol{z}_h) \frac{\delta \mathcal{H}_h(\boldsymbol{z}_h)}{\delta \boldsymbol{z}_h} ,
	\label{eq: poisson}
\end{align} 
where $\boldsymbol{z}_h=(\boldsymbol{u}_h,\varphi_h,B_h)$ and $\mathbb{A}$ is a skew symmetric operator
\reviewtwo{\cite{Eldred2019,BauerCotter2018}. }
The Poisson integrator proposed by \citet{Cohen2011} conserves energy and quadratic Casimirs for Poisson systems.

Let  $\mathcal{T}_{h}^t$ be a conforming partition of the temporal domain into  $N$ discrete intervals, $J^n \coloneqq (t^{n-1},t^{n}]$.
For simplicity, we assume all time intervals have equal length $\tau \coloneqq  T/N$. 
The discrete time nodes are $t^n \coloneqq n\tau$ for $n = 1,2, \ldots, N$.
For linear approximations in time, the Poisson integrator \cite{Cohen2011} is
\begin{align}
	\frac{ \boldsymbol{z}_h^{n} - \boldsymbol{z}_h^{n-1}}{\tau }
	=   \mathbb{A}\left( \frac{\boldsymbol{z}_h^n + \boldsymbol{z}_h^{n-1}}{2}  \right)  
	\int_{0}^1  \frac{\delta \mathcal{H}_h}{\delta \boldsymbol{z}_h}  \left( \boldsymbol{z}_h^{n-1} + s \left(  \boldsymbol{z}_h^{n} -  \boldsymbol{z}_h^{n-1} \right) \right) \mathrm{d}s,
	\label{eq: poisson integrator}
\end{align}
where the discrete variational derivatives of the Hamiltonian are exactly integrated in time \cite{BauerCotter2018,Eldred2019}. 
By construction of the Poisson integrator \cite{Cohen2011}, applying \cref{eq: poisson integrator} to the semi-discrete system \cref{eq: fem} yields temporal energy conservation. 
Unfortunately the chosen Poisson integrator \cite{Cohen2011} does not generalise to cubic Casimirs, such as discrete entropy, and we are not aware of any time integrator that preserves higher order Casimirs in general for non-canonical Hamiltonian systems.

\subsection{Temporal discretisation}

Motivated by the analysis of continuous entropy conservation in \cref{sec: system invariants} above,  we  represent the diagnostic buoyancy as a linear polynomial in time. 
This differs from the Poisson integrator of \citet{Cohen2011} where diagnostic variables are  constant in time.

We find temporal nodal values of $b_h^{n}\in \mathbb{V}_2$    using $\varphi_h^{n}, B_h^{n} \in \mathbb{V}_2$   in
the semi-discrete relation  \cref{eq: fem b semi}. 
Then, $b_h^{n-1/2} = (b_h^{n-1} + b_h^{n})/2$ is the mean value of a linear polynomial in time  for $t\in J^n$. 

Thus, the fully discrete system is: find $\boldsymbol{u}_h^{n} \in \mathbb{V}_{1}$, $\varphi_h^{n}, B_h^{n} \in \mathbb{V}_{2}$,  such that
\begin{subequations}
	\begin{align}
		\left( \boldsymbol{u}_h^{n} , \psi_{\boldsymbol{u}_h} \right) 
		- \left( \boldsymbol{u}_h^{n-1} , \psi_{\boldsymbol{u}_h} \right) 
		&+ \tau \left(  q_h^{n-1/2} ,\boldsymbol{F}_h^{\perp,n} \cdot \psi_{\boldsymbol{u}_h} \right) 
		- \tau	\left( \nabla  \cdot \psi_{\boldsymbol{u}_h}, \Phi_h^n \right)  \nonumber \\
		&- \tau g ( \psi_{\boldsymbol{u}_h},b_h^{n-1/2},\widetilde{b_h^n},\vartheta_h^n  )  
		- \tau s(\psi_{\boldsymbol{u}_h},b_h^{n-1/2},\vartheta_h^n)
		=0, \label{eq: fem u time}
		&\forall \psi_{\boldsymbol{u}_h} \in \mathbb{V}_{1} ,   \\
		\left( \varphi_h^n, \phi_{\varphi_h}\right) 
		-\left( \varphi_h^{n-1}, \phi_{\varphi_h}\right) 
		& + \tau \left( \nabla \cdot \boldsymbol{F}_h^n , \phi_{\varphi_h} \right) = 0, \label{eq: fem h time}   
		& 	\forall \phi_{\varphi_h} \in \mathbb{V}_{2} , \\
		\left(   B_h^n,  \phi_{B_h} \right)
		- 	\left(   B_h^{n-1},  \phi_{B_h} \right)
		&+\tau	g ( \boldsymbol{F}_h^n,b_h^{n-1/2},\widetilde{b_h^n},\phi_{B_h}  )   
		+\tau s(\boldsymbol{F}_h^n,b_h^{n-1/2},\phi_{B_h})
		=0,     \label{eq: fem B time}
		&\forall \phi_{B_h} \in \mathbb{V}_{2} ,
	\end{align} 
	\label{eq: fem time}
\end{subequations}
where $\widetilde{b_h^n} \in  \mathbb{V}_{2}$ satisfies,
\begin{align}
	\left(   \widetilde{b_h^n}b_h^{n-1/2},\widetilde{\phi}_h\right) &= \left( [b_hb_h]^{n-1/2} , \widetilde{\phi}_h\right) ,  \label{eq: b tilde time}
	& \forall \widetilde{\phi}_h \in \mathbb{V}_{2},
\end{align}
and $[b_h b_h]^{n-1/2}=( b_h^{n-1}b_h^{n-1} + b_h^n b_h^n  )/2$ 
is the mean value of the square of $b_h$. 
Quadratures up to the degree required are used to evaluate $\boldsymbol{F}^n_h \in \mathbb{V}_{1}$, $\Phi^n_h,\vartheta^n_h \in \mathbb{V}_{2}$, $q_h^{n-1/2} \in \mathbb{V}_{0}$   exactly in time using $\boldsymbol{u}_h^{n},\varphi_h^n,B_h^n$ in \cref{eq: fem diagnostic} \cite{Eldred2019,BauerCotter2018}.

\subsection{Temporal conservation}
\label{sec: temp conservation}

Discrete energy is conserved pointwise in time by construction of the Poisson integrator \cite{Cohen2011}. Taking $\psi_{\boldsymbol{u}_h} = \boldsymbol{F}^n_h$, $\phi_{\varphi_h} = \Phi_h^n$, $\phi_{B_h} = \vartheta_h^n$ in  \cref{eq: fem time} yields the temporal approximation of \cref{eq: dHdt} as
\begin{align}
	\reviewtwo{
	\int_{J^n}  \frac{\partial \mathcal{H}_h}{\partial t} 
	= \mathcal{H}_h^n - \mathcal{H}_h^{n-1}
	=\left( \boldsymbol{u}_h^{n} -\boldsymbol{u}_h^{n-1} ,  \boldsymbol{F}_h^n \right)  
	+\left( \varphi_h^n- \varphi_h^{n-1}, \Phi_h^n\right) 
	+ 	\left(   B_h^n- B_h^{n-1},  \vartheta_h^n \right)
	=0,        } 
	\label{eq: dHdt time}
\end{align}   
which is zero since skew symmetry is maintained. 
Proving  \cref{eq: fem h time} holds pointwise in space and time yields  discrete mass conservation.  
\begin{proposition}
	\label{prop: pointwise}
	The continuity equation \cref{eq: fem h time} holds pointwise in space and time.
	\begin{proof}
		Assuming continuity of time,
		\cref{eq: fem h} shows $ (\partial_t \varphi_h + \nabla \cdot \boldsymbol{F}_h) (x,\cdot) \in \mathbb{V}_2$ holds pointwise in space. 		
		Interpolation of the initial condition yields $\varphi_h^0\in \mathbb{V}_2$. Since $\varphi_h^n, \nabla \cdot \boldsymbol{F}_h^n \in \mathbb{V}_2$ for $n=1,2,\dots,N$, it is clear to see   
		$( \varphi_h^n - \varphi_h^{n-1} + \tau \nabla \cdot \boldsymbol{F}_h^n) \in \mathbb{V}_2$ holds pointwise in time.   
	\end{proof}
\end{proposition}

A key novelty of the fully discrete formulation in \cref{eq: fem time} is that the diagnostic buoyancy is represented as a continuous linear polynomial in time. 
Consequently, discrete entropy is a continuous cubic polynomial in time, which is not conserved pointwise in time by the chosen Poisson integrator \cite{Cohen2011}. 
We now prove the loss in exact entropy conservation depends on the accuracy of the temporal approximation. 
\begin{proposition}
	\label{prop: drift}
	Entropy conservation has a drift of size $\mathcal{O}(\tau)$, that is, $|\mathcal{S}_h^n - \mathcal{S}_h^0| = \mathcal{O}(\tau)$. 
	\begin{proof}
	First, observe the semi-discrete entropy, $\mathcal{S}_h (t)$, is conserved (\cref{prop: entropy conservation semi}), and thus equal to the initial entropy, $\mathcal{S}_h^0$, for all $t\in J$. 
    Second, recalling the order analysis for the time integration of the semi-discrete problem given  by \citet[Th. 4.3]{Cohen2011}, we obtain $\|b_h^n - b_h(t^n)\|_{L^\infty(\Omega)} + \| \varphi_h^n - \varphi_h(t^n)\|_{L^\infty(\Omega)} \leq \mathcal{O}(\tau)$.   
    Combining this with the definition of the fully discrete entropy yields
		\begin{subequations}
		\begin{align}
			\mathcal{S}_h^n &= \int_\Omega \frac{1}{2} \left[ \left( b_h^n - b_h\right) + b_h \right]
			\left[ \left( b_h^n - b_h\right) + b_h \right]
			\left[ \left( \varphi_h^n - \varphi_h\right) + \varphi_h  \right], \\
			&= \int_\Omega \frac{1}{2}b_h b_h \varphi_h + \int_\Omega \frac{1}{2}   \left( b_h^n - b_h\right)\left( b_h^n - b_h\right) \left( \varphi_h^n - \varphi_h\right)
			+ \dots \\
      &= \mathcal{S}_h^0 + \mathcal{O}(\tau).
		\end{align} 
		\end{subequations}
	Thus, $|\mathcal{S}_h^n - \mathcal{S}_h^0| = \mathcal{O}(\tau)$ and this  completes the proof. 
	\end{proof}
\end{proposition}

Subsequent to \cref{prop: drift}, \reviewtwo{it should be stated that} the relation in \cref{eq: dSdt semi} is not discretely preserved pointwise in time. 
To demonstrate the importance of \cref{eq: b tilde time} in regards to discrete entropy conservation,  we consider the temporal approximation of \cref{eq: dsdt semi 2} as  
\begin{align}
	\reviewtwo{
	\int_{J^n}  \frac{ \partial  \mathcal{S}_h  }{\partial t}
	=   \mathcal{S}_h^{n} -  \mathcal{S}_h^{n-1}  
	= \frac{1}{2} \left( [b_hb_h]^{n-1/2}, \nabla \cdot \boldsymbol{F}_h^n  \right) 	
	-   \frac{1}{2} \left(  \widetilde{b_h^n} b_h^{n-1/2}, \nabla \cdot \boldsymbol{F}_h^n \right) 
	- s_{\mathrm{up}}(\boldsymbol{F}_h^n,b_h^{n-1/2},b_h^{n-1/2}) . }
	\label{eq: dsdt time}
\end{align} 
Using $\widetilde{\phi}_h = \nabla \cdot \boldsymbol{F}_h^n$  in \cref{eq: b tilde time}
yields  
$\partial \mathcal{S}_h/\partial t = 0 $ when $\alpha = 0$, and
${\partial \mathcal{S}_h	}/{\partial t} < 0$ when $\alpha > 0 $.
We emphasise the choice of $\widetilde{b_h^n}$ means no spatial or temporal entropy conservation errors arise in the forcing terms. Thus, the only source of entropy conservation error is the approximation of temporal derivatives.
Numerical experiments in \cref{sec: results} below support this analysis.

\subsection{Constrained formulation}

Temporal conservation of Casimirs can be enforced using a global constraint.
In the context of discrete entropy conservation, we find  $b_h^n \in \mathbb{V}_2$, a minimiser of,
\begin{align}
	\mathcal{J}(b_h^n) &= \frac{1}{2}\left(b_h^n \varphi_h^n,b_h^n\right) - \left(B_h^n,b_h^n\right)  
	\quad 
	\text{such that }
	\quad  \mathcal{S}_h^n -\mathcal{S}_h^{n-1} = 0. 
	\label{eq: minimiser}
\end{align}
Equation 
\cref{eq: minimiser} can be written as:  find $\left[ b_h^n , \lambda  \right] \in \mathbb{V}_2 \times \mathbb{R}$ such that
\begin{equation}
	\begin{aligned}
	\left(	b_h^n \varphi_h^n, \phi_{b_h} \right)
	- \left(  B_h^n , \phi_{b_h} \right)
	+ \lambda \left( b_h^n \varphi_h^n,\phi_{b_h} \right)
	+ \frac{1}{2}\mu \left( 	b_h^n	 \varphi_h^n ,b_h^n\right)
	&=  \mu\mathcal{S}_h^{n-1}, 
	&\forall \left[\phi_{b_h},\mu \right]\in \mathbb{V}_2 \times \mathbb{R},
	\label{eq: fem b constrained}  
	\end{aligned}
\end{equation}
where  $\lambda \in \mathbb{R}$ is a Lagrange multiplier. 
Diagnosing $b_h^n$ using \cref{eq: fem b constrained} instead of \cref{eq: fem b semi} means the analysis of  temporal entropy conservation in \cref{sec: temp conservation} 
is trivial,
since we explicitly enforce $\tau \partial \mathcal{S}_h/\partial t=\mathcal{S}_h^n -\mathcal{S}_h^{n-1} = 0$  pointwise in time.  
It follows from \cref{prop: drift} that $\lambda$ is sufficiently small and corrects an $\mathcal{O}(\tau)$ drift in exact entropy conservation. 
Further,  using \cref{eq: fem b constrained} perturbs $b_h^n$ everywhere  in \cref{eq: fem time}. Since skew symmetry is maintained independent of $b_h^n$, discrete energy conservation holds as in \cref{eq: dHdt time} by construction of the chosen Poisson integrator \cite{Cohen2011}. 

\subsection{Quasi-Newton approach}
\label{sec: quasi-newton}

The fully discrete system  \cref{eq: fem time} is an implicit   non-linear problem, for which using a quasi-Newton approach to approximate the Jacobian can reduce  computational expense.  
The Jacobian is computed once per time step as 
\begin{equation}
	\begin{aligned}
		\langle  \mathbf{J}( \boldsymbol{\delta u_h},\delta \varphi_h, \delta B_h), (\psi_{\boldsymbol{u}_h},\phi_{\varphi_h},\phi_{B_h}) \rangle &= 
		\left(    \boldsymbol{\delta u}_h,  \psi_{\boldsymbol{u}_h} \right)
		+ \frac{\tau}{2} \left(  \omega_h^{n-1} , \boldsymbol{\delta u}_h^\perp \cdot \psi_{\boldsymbol{u}_h} \right)\\
		&	- \frac{\tau}{2} \left( \nabla \cdot \psi_{\boldsymbol{u}_h}, \frac{\delta B_h}{2} \right) 
		- \frac{\tau}{2}  \left( \nabla \cdot \psi_{\boldsymbol{u}_h},  b_h^{n-1} \frac{\delta \varphi_h}{2}  \right)  \\
		&+ \left(  \delta \varphi_h ,\phi_{\varphi_h} \right)
		+\frac{\tau}{2}  \left(   \nabla \cdot \boldsymbol{ \delta u}_h, \varphi_h^{n-1} \phi_{\varphi_h}  \right) \\
		&+\left( \delta B_h, \phi_{B_h} \right)
		+ \frac{\tau}{2} \left(  \nabla \cdot  \boldsymbol{\delta u}_h,  b_h^{n-1} \varphi_h^{n-1} \phi_{B_h} \right) ,
	\end{aligned}
	\label{eq: jacobian}
\end{equation}
where $\omega_h^{n-1}\in \mathbb{X}_{01}$ is diagnosed by  setting $\omega_h^{n-1}=q_h^{n-1}\varphi_h^{n-1}$ in \cref{eq: fem q}.   
Equation 
\cref{eq: jacobian}
originates from an implicit midpoint discretisation of
the linear thermal shallow water equations, in which fast linear waves  are accounted for and non-linear coupling terms are omitted \cite{Eldred2019,Lee2022,BauerCotter2018,Wimmer2020}.  
The linearisation in the current study differs from other works \cite{Eldred2019,Lee2022} as we consider perturbations from the solution at $t^{n-1}$ 
by using $\varphi_h^{n-1},b_h^{n-1}$ in \cref{eq: jacobian}
instead of the initial usual mean flow state. In doing so, we account for turbulent states which significantly deviate from the resting steady state as simulations progress. 
Including upwinded numerical fluxes may result in solver convergence issues when simulating a mature turbulent state (see  \cref{sec: results thermal instability}). 
In these situations, the  Jacobian is recomputed at every non-linear iteration and includes approximated centred and upwinded fluxes.

\subsection{Implementation remarks}
The quasi-Newton approach is implemented using Gridap.jl \cite{Badia2020,Verdugo2022}, a Julia package for finite element computations. 
The   implicit time stepping framework within Gridap.jl   is modified
to solve the linear diagnostic problem prior to evaluating the residual of the non-linear prognostic problem \cite{Eldred2019}. 
The linear diagnostic problem is constructed as explicit. The current guesses for $\boldsymbol{u}_h^n,\varphi_h^n,B_h^n$ are used to:
\begin{enumerate}
	\item Evaluate $\boldsymbol{F}^n_h \in \mathbb{V}_{1}$, $\Phi^n_h,\vartheta^n_h, b_h^n \in \mathbb{V}_{2}$, $q_h^{n-1/2} \in \mathbb{V}_{0}$   exactly in time via \cref{eq: fem diagnostic}.
	\item Construct $b_h^{n-1/2}$ and $[b_h b_h]^{n-1/2}$, and diagnose $\widetilde{b_h^n}$  via \cref{eq: b tilde time}.
\end{enumerate}
We use GridapDistributed.jl \cite{Badia2022}, GridapPETSc.jl \cite{GridapPETSc} and GridapSolvers.jl \cite{GridapSolvers}, to conduct tests on the Gadi@NCI Australian supercomputer.  
The source code for this study is  available on Zenodo \cite{Zenodo}.  

Like other upwinded schemes \cite{Brezzi2004}, the upwinding parameter $\alpha$ depends on the direction of flow, and requires a numerical implementation of the signum function, $\sign(x)$.
One possibility is  
\begin{align}
	\sign(x) = 
	\begin{cases}
		1& x > \epsilon, \\
		- 1&x < -\epsilon, \\
		0 & \text{otherwise},
	\end{cases}
	\label{eq: sign x}
\end{align}
which converges to $\sign(x)$ as $\epsilon \rightarrow 0 $ for $\epsilon \geq 0$. 
However \cref{eq: sign x} is a non-differentiable function that may   cause issues in numerical solvers \cite{Badia2017}. 
An alternative approach is to approximate $\sign(x)$ by a smooth function \cite{Badia2017} such as 
\begin{align}
	\sign(x) = \frac{x}{\sqrt{x^2 + \epsilon^2}} , \label{eq: sign x smooth}
\end{align}
which also converges to $\sign(x)$ as $\epsilon \rightarrow 0 $.
We refer to \cref{eq: sign x,eq: sign x smooth} as the \textit{hard} and \textit{soft} signum functions respectively, and explore their effects in the numerical experiments below.
The value of $\epsilon$  is  based on values of $||\boldsymbol{F}_h^n\cdot \boldsymbol{n}^+||$ from experiments using the centred scheme.

%% file: results.tex
\section{Numerical experiments}
\label{sec: results}

To examine the ability of the new finite element approximation  to conserve system invariants over long periods of time, we first evaluate convergence under $h$-$p$ refinement and then consider transient test cases that exhibit turbulent flows.

\subsection{Convergence test}
\label{sec: convergence test}
Convergence is assessed using a steady zonal thermogeostrophic balance test \cite{Eldred2019}. 
In all simulations, the time step is $\tau = C  h$ where $C = \mathrm{CFL}/(p^2\sqrt{g\varphi_0})$, $p$ is the spatial  polynomial order,  $h$ is the spatial step, 
and $\sqrt{g\varphi_0}$ represents the mean speed of gravity waves for which pressure gradient forces dominate \cite{Vallis2006}.  
Simulations are conducted in a reference frame where the domain length is an order 1 quantity. This is generated by rescaling the initial condition using the length scale $L_0$, time scale $T_0$,  velocity scale $U_0$, depth scale $H_0$, and buoyancy scale $b_0$.  
This yields the Rossby number $\mathrm{Ro} = U_0/(T_0L_0)$  and the Burgers number $\mathrm{Bu} =b_0 H_0 / (T_0^2 L_0^2)$.

The rescaled initial condition is
\begin{align}
	{u}_1 &= \cos ({x}_2), & 
	{u}_2 &= 0, &
	{\varphi} &= 1 - \frac{\mathrm{Ro} }{\mathrm{Bu} }\sin ({x}_2), &
	{b} &= 1 + c {\varphi}, 
\end{align}
where  ${\boldsymbol{u}}=(u_1,u_2)$, ${B} = {b}{\varphi}$, ${x}_{1,2}$ are the horizontal and vertical spatial coordinates  respectively, and $c=0.05$. 
In the rescaled frame, ${\Omega} = [0,2\pi]^2$ is doubly periodic, ${g}=\varphi_0 = 1$, ${f} = \mathrm{Ro}/\mathrm{Bu}$, and the duration corresponds to  5 days \cite{Ricardo2024-dg}. 
The scaling parameters are: $L_0 = 6371120 \text{ m}$,  $T_0 = 6.147\times 10^{-5} \text{ s}^{-1}$, $U_0 = 20 \text{ m s}^{-1}$, $H_0 = 5960 \text{ m}$, and $b_0 = 9.80616 \text{ m s}^{-1}$.

\begin{figure}[h!]
	\centering 
	\includegraphics[width=\textwidth]{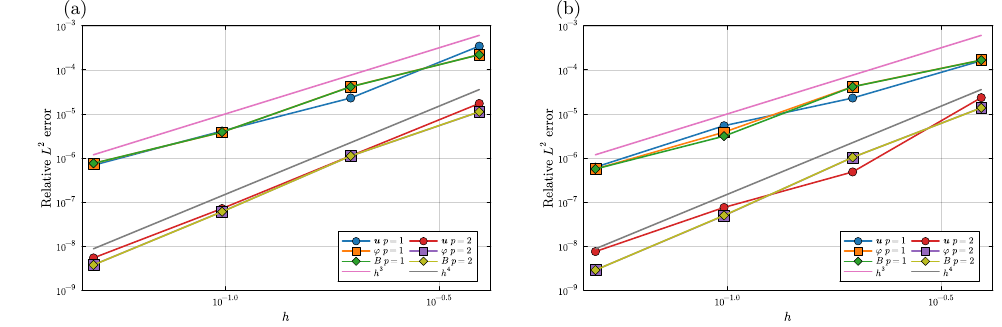} 
	\caption{Convergence  of the zonal thermogeostrophic test case showing  relative $L^2$ error between initial  and final solutions for (a) the centred scheme and (b) the upwinded scheme (hard signum, $\epsilon=10^{-4}$), with 16, 32, 64, 128 spatial elements in each case. 
	}
	\label{fig: convergence}
\end{figure}

Convergence of the centred scheme is tested using the relative $L^2$ norm between the initial and final solutions in the absence of stabilisation \cite{Eldred2019}. 
\cref{fig: convergence}(a) shows third and fourth order convergence for \reviewtwo{$p=1$ and $p=2$ finite elements respectively}, for a range of spatial and temporal resolutions.
\reviewtwo{The apparent super-convergence could be attributed to the fact that this test case is an extremely smooth test case, where no turbulent flow structures develop. }
For each line in \cref{fig: convergence}, ${h}$ and ${\tau}$ reduce simultaneously at rate $C=\mathrm{CFL}/p^2$,  where  $\mathrm{CFL}=0.2$ for $p=1$ and $\mathrm{CFL}=0.1$ for $p=2$. 
Similar convergence rates are observed in \cref{fig: convergence}(b) for the upwinded scheme.   

\subsection{Case study 1: thermal instability}

\label{sec: results thermal instability}

We now consider transient case studies 
to confirm adequate conservation of system invariants under turbulent flow conditions and over long time periods relative to the onset of geostrophic turbulence. 
Discrete conservation is demonstrated using normalised values of discrete energy \cref{eq: discrete energy t}, entropy \cref{eq: discrete entropy semi} and mass ($\mathcal{M}_h(t) = \int_{\Omega} \varphi_h $) on a logarithmic scale over time. 
The change in entropy due to forcing terms, $\partial \mathcal{S}_h/\partial t$ in \cref{eq: dsdt time}, is also shown. 
\reviewtwo{The following numerical experiments correspond to $p=1$ finite elements and $\mathrm{CFL}=0.2$. Based on the convergence results evident in \cref{fig: convergence}, using $p=2$ finite elements would require a smaller CFL value. 
}

\reviewone{The analysis of conservation of system invariants in \cref{sec: temp conservation,sec: conservation} is based on a converged non-linear solution. 
We recognise other models solve with a fixed number of non-linear iterations per time step, as opposed to converging to a chosen tolerance \cite{Maynard2020,Lee2024,Wood2013,Lee2024_helmholtz}. Such studies may also utilise preconditioning and multigrid methods to accelerate convergence \cite{Maynard2020}. 
In the present study, the intent of the following numerical experiments is to demonstrate   conservation of systems invariants in the presence of highly turbulent flow.}
So we accept non-linear solutions that converge to a tolerance of $10^{-12}$, meaning conservation errors up to $\mathcal{O}(10^{-12})$ are reasonable.  
\reviewonee{
A maximum of 50 non-linear iterations per time step is used in all simulations, even though the centred scheme generally converges in less than 20 non-linear iterations per time step}.

The first transient case study involves a thermal instability \cite{Eldred2019,Gouzien2017,Zeitlin2018,Kurganov2021}. 
Other studies simulate this test case using an upwinded finite volume scheme  \cite{Kurganov2021,Gouzien2017,Zeitlin2018}. A previous finite element study shows energy conservation at early times \cite{Eldred2019}, but entropy conservation, the evolution of complex non-linear flows, and the effect of upwinding is not considered.
We extend these works using the novel centred and upwinded fluxes to simulate the thermal instability test case \reviewtwo{over a} long time period while examining conservation of system invariants.

The initial condition is 
\begin{equation}
\begin{aligned}
	{u}_1 &= -U_0r  \exp\left(\frac{1-r^\beta}{\beta}\right) \sin(\phi) + \varepsilon, &
	{u}_2 &= 	U_0r  \exp\left(\frac{1-r^\beta}{\beta}\right)\cos(\phi)  + \varepsilon ,  \\
	{\varphi} &= 1 - \varepsilon , &
	{b} &= 1 - 2 \frac{ \mathrm{Ro} }{\mathrm{Bu} } \left( 
	\exp\left(\frac{1-r^2}{2}\right) + \frac{\mathrm{Ro} }{2} \exp\left(1-r^2\right)
	  \right) + \varepsilon  ,
\end{aligned}
\end{equation}

\noindent
where  
\begin{align}
	\varepsilon &=  0.01\exp\left(-60(r-r_c)^2 \right)\sin \left( 6\pi (r-r_c)\right) \cos(4\phi ), \label{eq: pertubation}
\end{align}

\noindent
is a perturbation, 
${r}^2 = {x}_1^2 + {x}_2^2$ is the radial distance, $\tan (\phi) = x_2/x_1$ is the polar angle, and $\Omega = [-4,4]^2$ is doubly periodic.
The duration, $T = 100$, is sufficiently long for a mature turbulent state to develop.
The parameters are:
$\beta  = 2$, $r_c = 0.5$, $g = \varphi_0 =  f = L_0 = H_0 = b_0 = T_0= \mathrm{Bu} = 1$, and $U_0 = \mathrm{Ro} = 0.1$.

\begin{figure}[h!]
	\centering 
	\includegraphics[ 
	width=0.65\textwidth, keepaspectratio
	]{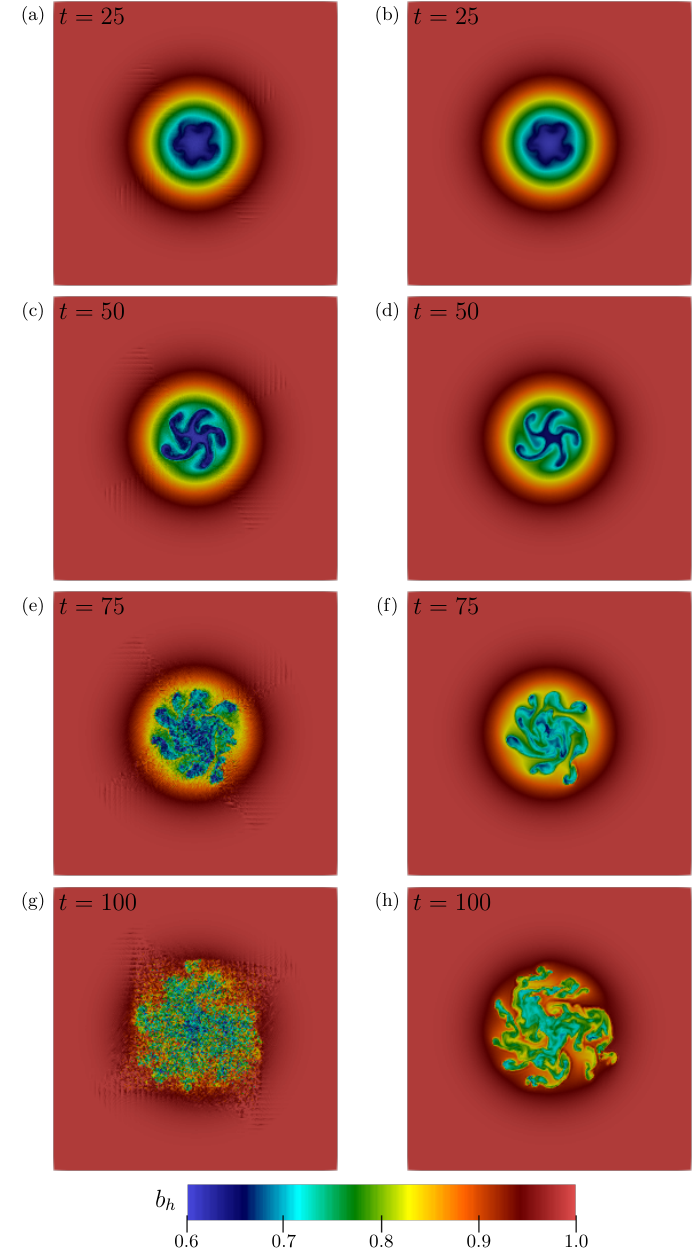} 
	\caption{ 
		Evolution of buoyancy $b_h$ for the thermal instability case study where (a,c,e,g) relate to the centred scheme and (b,d,f,h) correspond to the upwinded scheme (hard signum, $\epsilon = 10^{-4}$). Snapshots are shown at $t=25,50,75,100$ where time is unit-less.
		Parameters: $n_s=192$ spatial elements, $p=1$ spatial finite elements, $\mathrm{CFL} = 0.2$.
	}
	\label{fig: instability profiles}
\end{figure}

\cref{fig: instability profiles} shows the evolution of the thermal instability for the centred  and upwinded schemes. 
Similar to other studies that also consider this test case \cite{Eldred2019,Gouzien2017}, the initial growth evident in \cref{fig: instability profiles}(a)--(d) reflects the wave number of 4 used in  \cref{eq: pertubation}, 
and non-linear saturation of the instability occurs as the simulation progresses (\cref{fig: instability profiles}(e)--(h)). 
Comparing the buoyancy field at $t=25$ (\cref{fig: instability profiles}(a,b)) that obtained at $t=100$ (\cref{fig: instability profiles}(g,h)) shows a large variation in buoyancy. Regardless, both the centred and upwinded schemes resolve the highly non-linear flow that develops as the simulation progresses. 
This is a direct consequence of the new linearised Jacobian \cref{eq: jacobian}, which allows for robust convergence for both centred and upwinded fluxes in the presence of significant variations from the initial mean flow state. 
As expected, including upwinded fluxes (\cref{fig: instability profiles}(b,d,f,g)) suppresses spurious oscillations which arise in the absence of stabilisation (\cref{fig: instability profiles}(a,c,e,g)). 
\reviewtwo{Therefore, we recommend using the upwinded scheme from the perspective of obtaining smooth solutions. }

\begin{figure}[h!] 
	\includegraphics[width=\textwidth]{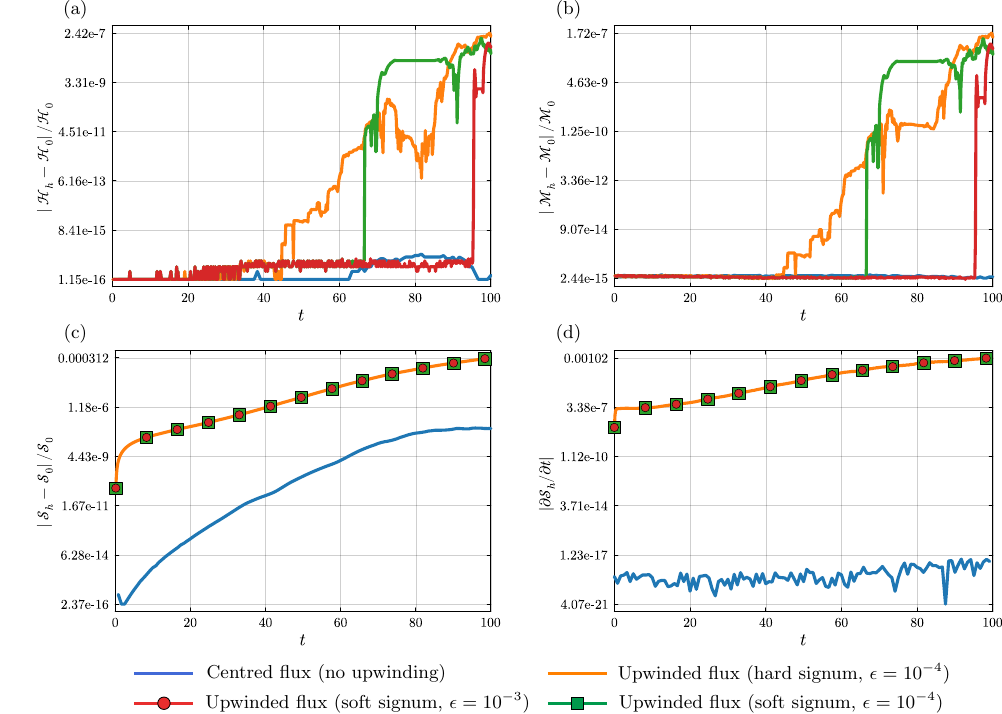} 
	\caption{Conservation errors
		for the thermal instability case study.
		(a,b,c) show  normalised values of energy, mass and entropy, and (d) illustrates the change in entropy due to forcing terms \cref{eq: dsdt semi 2}.
		Each figure compares the centred and upwinded schemes for different approximations of  signum and $\epsilon$ values.  The vertical axis is  logarithmic  and the horizontal axis shows unit-less time.  
		Parameters: $n_s=192$ spatial elements, $p=1$ spatial finite elements, $\mathrm{CFL} = 0.2$.  
	}
	\label{fig: instability conservation}
\end{figure}

We now assess the ability of the centred and upwinded schemes to conserve system invariants in the presence of turbulent flow. 
In the absence of stabilisation, 
\cref{fig: instability conservation}(a,b) shows energy and mass  
conservation, while  \cref{fig: instability conservation}(c) demonstrates the entropy conservation error 
increases to $\mathcal{O}(10^{-8})$ over time even though ${\partial \mathcal{S}_h}/{\partial t}$ is machine zero for the entire duration (\cref{fig: instability conservation}(d)). 
This supports the analysis in \cref{sec: temp conservation} above that entropy is not conserved pointwise in time. 
While an entropy conservation error of $\mathcal{O}(10^{-8})$ may seem large relative to energy and mass, it is comparable with  other studies that consider temporal conservation of quadratic tracer invariants \cite{Lee2022}. 
Thus, we conclude the centred scheme stably resolves turbulent dynamics over long periods of time and conserves system invariants.

For the upwinded scheme, the spikes in energy and mass evident in \cref{fig: instability conservation}(a,b) correspond to instances when  the non-linear solver fails to converge. 
\reviewone{ In general, conservation errors obtained from numerical experiments are bounded by the tolerance of the non-linear solver. 
Consequently, non-linear solutions that fail to converge can lead to an increase in conservation error. 
As discussed above, we accept conservation errors up to $\mathcal{O}(10^{-12})$.  
The results in \cref{fig: instability conservation}(a,b) demonstrate that failing to converge to the chosen  non-linear tolerance leads to increased conservation errors of up to $\mathcal{O}(10^{-7})$ in energy and mass. 
While  conservation errors of $\mathcal{O}(10^{-7})$ may seem large, such errors are comparable with other studies that choose to accept non-linear solutions after a fixed number of non-linear iterations per time step \cite[Fig. 5]{Lee2024_helmholtz}.}

Values of $\epsilon=10^{-3}$ and $\epsilon=10^{-4}$ are chosen to approximate $\sign(x)$ in the upwinded scheme since $||\boldsymbol{F}_h^n\cdot \boldsymbol{n}^+||\approx 0.1$ for the centred scheme.  
When using the hard signum function with $\epsilon = 10^{-4}$, solver convergence issues arise at $t\approx 45$. 
In contrast, using the soft signum function for $\epsilon = 10^{-4}$ and  $\epsilon=10^{-3}$ delays the onset of solver convergence issues to $t\approx 70$ and $t\approx 98$ respectively. 
Therefore, using the soft signum function and increasing $\epsilon$ assists solver convergence as turbulent flow dynamics evolve.

A consequence of turbulent flow dynamics is that the direction of flow may change within discrete time intervals. 
Since the Jacobian is recomputed at every non-linear iteration and includes both approximated centred and upwinded fluxes, we make every effort to minimise solver convergence issues regardless of the choice of signum function or $\epsilon$ value. 
As with any numerical simulation of partial differential equations, using an increasingly fine computational mesh and  sophisticated non-linear solver  can reduce convergence issues.
We acknowledge non-linear solver issues may arise given the exceedingly small numbers  required to demonstrate conservation  in the context of highly turbulent flow. 

The logarithmic axis in \cref{fig: instability conservation}(c,d) helps exemplify multiple orders of magnitude difference in entropy conservation between the centred and upwinded schemes. 
Using a linear vertical axis to represent conservation error shows that indeed ${\partial \mathcal{S}_h}/{\partial t}<0$ for the upwinded scheme, and entropy is monotonically damped (not shown). 
This supports the analysis of temporal entropy conservation in \cref{sec: temp conservation}.

\subsection{Case study 2: double vortex}
\label{sec: results merging}

The second transient case study describes a double vortex \cite{Eldred2019}. 
Test cases involving vortex pair interactions are often used to assess numerical schemes for the rotating shallow water equations \cite{Giorgetta2009,Reich2006,Mcrae2013,Staniforth2006}. 
A previous study adapted such a test case to the thermal shallow water equations by considering the time evolution of two vortices interacting in a zonally varying buoyancy field \cite{Eldred2019}. 
While this previous study considered an energy conserving finite element formulation \cite{Eldred2019}, entropy conservation was not of concern. 
Thus, we use the novel centred fluxes proposed in the current study to achieve semi-discrete entropy conservation, and demonstrate the use of Lagrange multipliers to rectify small losses in exact entropy conservation.

The rescaled initial condition is
\begin{equation}
	\begin{aligned}
		{u}_1 &= -
		\eta_{21}\varepsilon_1 - \eta_{22} \varepsilon_2   , &
		{u}_2&= \eta_{11}\varepsilon_1 + \eta_{12} \varepsilon_2  , \\
		{\varphi} &= 1 -  \varphi_c \left( \varepsilon_1 + \varepsilon_2  - 4 \pi \sigma^2 
		\right) ,&
		{b} &=   1 + c\sin \left(  2 \pi  {x}_1-\pi   \right)  , 
	\end{aligned}
\end{equation}
where   
\begin{align}
	\varepsilon_j &= \exp\left( -\frac{1}{2} (\gamma_{1j}^2 + \gamma_{2j}^2)  \right), &  
	\gamma_{ij}   &= \frac{1}{\pi \sigma } \sin \left( \pi ({x}_i-{c}_j) \right), &
	\eta_{ij} &= \frac{1}{2\pi \sigma  } \sin \left(  2\pi  ({x}_i-{c}_j) \right),  
\end{align}
for $i,j=1,2$,  $\varphi_c = h_0/H_0$, $\sigma = 3/40$, ${c}_1 = 0.4$,  ${c}_2 = 0.6$ and $c=0.05$ \cite{Eldred2019}.
In the rescaled frame, ${\Omega} = [0,1]^2$ is doubly periodic,  ${g} = {\varphi}_0 = 1$, and  ${f}$  is computed such that $\mathrm{Bu}$ is constant between the rescaled and unscaled frames. 
The duration is chosen to allow mature turbulent structures to develop.  
The scaling parameters are: $L_0 = 5 \times 10^6$ m, $T_0 = 6.147 \times 10^{-5}$ s$^{-1}$, $h_0 = 75 \text{ m}$, $H_0 = 750 \text{ m}$,  $b_0 = 9.80616 \text{ m s}^{-2}$ 
 and $U_0 = (b_0  h_0)/(T_0L_0\sigma  )  $    \cite{Eldred2019}.

\begin{figure}[h!]
	\centering 
	\includegraphics[width=0.65\textwidth]{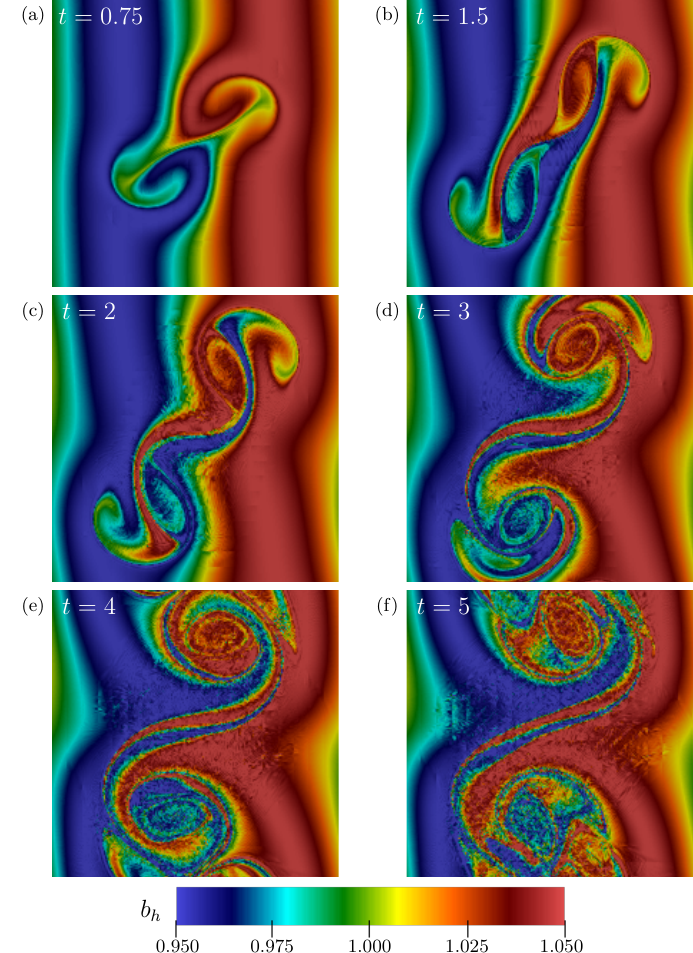} 
	\caption{ 
		Evolution of buoyancy $b_h$ for the double vortex case study, simulated with the centred scheme. Snapshots are shown at $t=0.75,1.5,2,3,4,5$ where $t$ is unit-less.  Parameters: $n_s=192$ spatial elements, $p=1$ spatial finite elements, $\mathrm{CFL} = 0.2$.
	}
	\label{fig: merging profiles}
\end{figure}

\cref{fig: merging profiles} shows the centred scheme adequately resolves turbulent flow structures over long periods of time with minimal grid scale noise.
Similar to other studies that consider vortex pair interactions
\cite{Giorgetta2009,Reich2006,Mcrae2013,Staniforth2006,Eldred2019}, 
we observe the formation of two distinct vortices (\cref{fig: merging profiles}(a)) which collide as a mature turbulent state is reached (\cref{fig: merging profiles}(f)). 
The ability of the centred scheme to resolve such complex features is further evidence that the new linearised Jacobian \cref{eq: jacobian} provides robust convergence with respect to large variations from the initial mean flow. 

\begin{figure}[h!]
	\centering 
	\includegraphics[width=\textwidth]{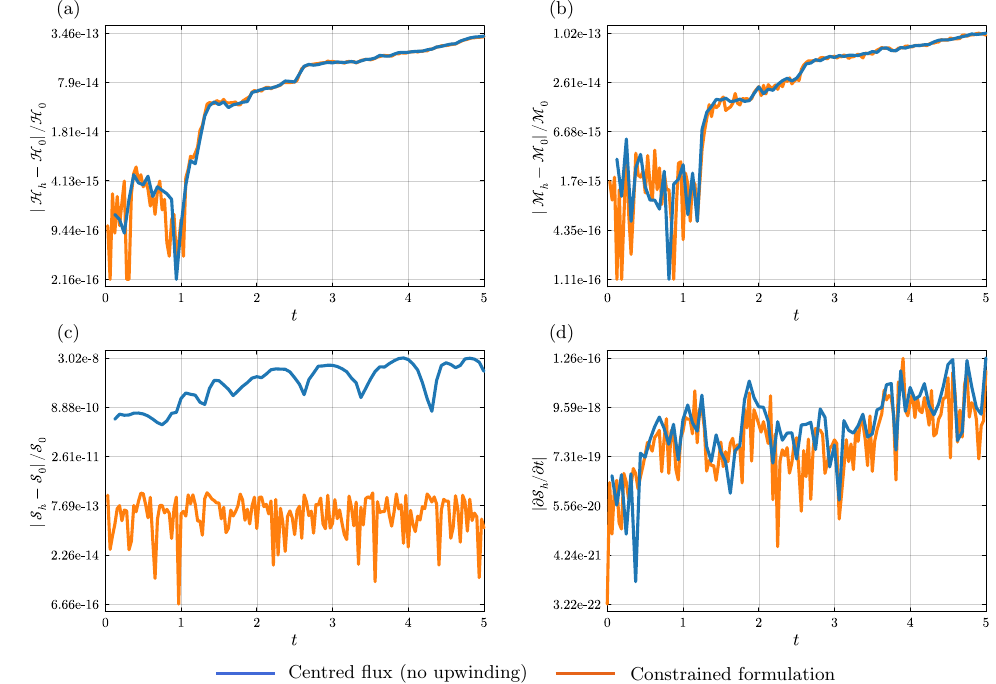} 
	\caption{ 
		Conservation errors for the double vortex case study. 
		(a,b,c)  show  normalised values of energy, mass and entropy, and (d) illustrates the  change in entropy due to forcing terms \cref{eq: dsdt semi 2}. Each figure compares the centred  (blue) and constrained schemes (orange). 
		The vertical axis is logarithmic and the horizontal axis shows unit-less time. Parameters: $n_s=64$ spatial elements, $p=1$ spatial finite elements, $\mathrm{CFL} = 0.2$. 	
	}
	\label{fig: merging conservation}
\end{figure}

For the centred scheme, 
\cref{fig: merging conservation}(a,b) shows conservation of  energy and  mass to machine precision. 
Similar to \cref{sec: results thermal instability}, \cref{fig: merging conservation}(c) shows a drift in entropy conservation from machine precision to $\mathcal{O}(10^{-8})$ as the simulation progresses even though  $\partial \mathcal{S}_h/\partial t$ is machine zero (\cref{fig: merging conservation}(d)). 
Thus, we conclude  entropy is not conserved pointwise in time. 

To achieve exact entropy conservation, we use the constrained formulation in \cref{eq: fem b constrained}, for which entropy conservation is explicitly enforced using a Lagrange multiplier. 
Conservation errors in energy, mass and  $\partial \mathcal{S}_h/\partial t$ \reviewtwo{are} comparable between the centred and constrained formulations (\cref{fig: merging conservation}(a,b,d)), and 
\cref{fig: merging conservation}(c) shows entropy is conserved within the tolerance of the non-linear solver. 
\reviewone{ In our experience, solving \cref{eq: fem b constrained} only requires a few more non-linear iterations relative to the unconstrained formulation. 
}
Thus, in situations where exact entropy conservation is required, we advise use of the constrained formulation.


%% file: conclusion.tex
\section{Conclusion and future work}
\label{sec: conclusion}

In this study, we develop a new finite element discretisation for the thermal shallow water equations. 
Novel centred and upwinded fluxes are developed to include   a specific formulation of buoyancy, $\widetilde{b_h}$, at the semi-discrete level.
We  prove the centred fluxes are energy and entropy conserving, while the upwinded fluxes are energy conserving and entropy damping. 
Using a Poisson time integrator \cite{Cohen2011} and an original  temporal representation of  discrete thermodynamic quantities, we construct a fully discrete system that conserves entropy up to the accuracy of the temporal approximation, and conserves energy by construction. 
In particular, the temporal formulation of $\widetilde{b_h}$ ensures the forcing terms conserve entropy in time, such that temporal conservation errors only arise from approximating time derivatives. 
The developed quasi-Newton approach utilises the solution at progressive time levels instead of the typical mean flow state \cite{Eldred2019,Lee2022} to capture the evolution of turbulent dynamics.
Transient case studies involving a thermal instability \cite{Eldred2019,Gouzien2017,Zeitlin2018,Kurganov2021} and double vortex \cite{Eldred2019,Giorgetta2009} 
illustrate the ability of the novel centred and upwinded fluxes, and linearised Jacobian to  stably resolve complex flow structures over long periods of time.

A consequence of the chosen Poisson time integrator \cite{Cohen2011} is that cubic Casimirs are not preserved for non-canonical Hamiltonian systems. 
Thus, discrete entropy is not conserved pointwise in time. 
To address this, we propose a constrained formulation using Lagrange multipliers  which numerical experiments confirm corrects small temporal losses in exact entropy conservation. 
\reviewtwo{
The constrained formulation is recommended in situations where exact entropy conservation is required. 
However, the unconstrained upwinded scheme, for which entropy is monotonically damped, is generally preferred to smooth spurious oscillations that arise in the absence of stabilisation.
From the analysis and results presented in this study, we have identified the need to develop a Poisson time integrator that preserves higher order Casimirs for Poisson systems in general.}
While Poisson time integrators  that preserve higher order Casimirs for canonical Hamiltonian systems have recently been developed  \cite{Modin2020}, the extension to non-canonical Hamiltonian systems is an open problem.

There are many potential avenues to extend the framework presented in this study. 
We take a fundamental approach and showcase the new finite element scheme using planar test cases. 
One avenue for extension is to simulate spherical test cases  \cite{Ricardo2024-dg}. 
This is feasible since the numerical method and analysis presented in this study does not depend on the choice of coordinate system. 
The  quasi-Newton approach and linearised Jacobian developed in this study  could be extended to other problems that consider quasi-Newton methods in conjunction with multigrid preconditioners for atmospheric models \cite{Maynard2020}.
We always consider skew symmetric formulations  of the thermal shallow water equations. 
So the developed numerical schemes and analysis could be interpreted in the context of the compressible Euler equations \cite{Ricardo2023}. 
This would require careful consideration of the spatial and temporal approximation of relevant thermodynamic quantities and variational derivatives, and is left for future consideration.